\documentclass[prb,reprint,showpacs,twocolumn,superscriptaddress]{revtex4-2}

\usepackage[utf8]{inputenc}
\usepackage[T1]{fontenc}
\usepackage{graphicx}
\usepackage{dcolumn}
\usepackage{bm}
\usepackage{amsmath}
\usepackage{color}
\usepackage{gchords}
\usepackage{xcolor}
\usepackage{ulem}

\newcommand{\bra}[1]{\left\langle #1 \right|}
\newcommand{\ket}[1]{\left| #1 \right\rangle}

\newcommand{\ketbra}[2]{\left|#1\middle\rangle\middle\langle#2\right|}

\newcommand{\ad}[1]{\hat a^{\dagger}_{#1}}

\newcommand{\UFSC}{Departamento de Física, Universidade Federal de Santa Catarina,
88040-900, Florianópolis, SC, Brazil}

\newcommand{\UFMG}{Departamento de F\'{\i}sica - ICEx - Universidade Federal de Minas Gerais,
Av. Pres. Ant\^onio Carlos 6627 - Belo Horizonte - MG - Brazil - 31270-901.}

\newcommand{\UFF}{Instituto de Física, Universidade Federal Fluminense, 24210-346 Niter\'oi, Brazil.}

\begin{document}

\title{Quantum correlations, entanglement spectrum and coherence of two-particle reduced density matrix in the Extended Hubbard Model}

\author{Diego L. B. Ferreira}
\email{diegobragaferreira@gmail.com}
\affiliation{\UFMG}
\author{Thiago O. Maciel}
\affiliation{\UFSC}
\author{Reinaldo O. Vianna}
\affiliation{\UFMG}
\author{Fernando Iemini}
\affiliation{\UFF}

\date{\today}

\begin{abstract} 

We study the ground state properties of the one-dimensional extended Hubbard model at half-filling from the perspective of its \textit{particle} reduced density matrix. 
We focus on the reduced density matrix of $2$ fermions and perform an analysis of its quantum correlations and coherence along the  different phases of the model. Specifically, we study its (i) entanglement entropy, (ii) $\ell_1$ norm of coherence, (iii) irreducible two-body cumulant matrix and (iv) entanglement spectrum. 
Our results show that these different properties are complementary to each other depending on the phase of the system, exhibiting peculiar behaviors such as discontinuities, maximum or minimum values at the quantum phase transitions, thus providing a qualitative view of the phase diagram of the model.
In particular, in the superconducting region, we obtain that the entanglement spectrum signals a transition between a dominant singlet (SS) to triplet (TS) pairing ordering in the system. 
Moreover, from the analysis of the dominant eigenvector in the reduced state, we can relate the SS-TS transition to the spatial separation between the fermion pairs in the two different pairing orderings.
The entanglement gap is also able to 
highlight a transition - at a few-body level - in the groundstate wavefunction, not discussed previously in the literature. While other quantifiers are less sensitive to few-body defects in the wavefunction, the entanglement gap can work as a magnyfying glass for these, capturing such small fluctuations.
\end{abstract}

\maketitle

\section{Introduction}

A wide range of condensed matter phases can emerge when the constituents of the system are brought together and allowed to interact with each other. 
Due to the many-body interactions among its contituents, when the number of constituents is large, the system condenses into a collective behavior with specific macroscopic properties \cite{Fraser2006}.
Most familiar examples include magnetism, arising from the exchange interaction between local magnetic moments; solids and liquids, which arise from the electromagnetic forces between atoms; superconductors or superfluids, arising from the interaction between fermions or bosons; as well as more unconventional ones such as topological phases, emerging from nonlocal correlations among its constituents \cite{Xiao2013}.

There are different ways to analyse many-body phases.  
In conventional phases one can usually rely on the structure of its local correlations, and corresponding order parameters, which provide information about the macroscopic properties of the system. 
In recent years, different approaches have also been put forward, relying on interesting connections between Quantum Information and Condensed Matter theories. 
Much activity at the border of these fields and many interesting concepts have been addressed \cite{amico08}. 
In particular, quantum information insights about many-body entanglement has proved a powerful tool in order to study and characterize many-body systems, giving a unique perspective in our understanding of condensed matter phases \cite{BeiZeng2018}. 

The entanglement between the constituents of the system is shown to be tighly connected to the characteristics of the different phases a model can support. 
When the system is driven along a quantum phase transition, the entanglement is expected to show peculiar critical behaviors, allowing for a qualitative display of the transition and a deeper characterization of the many-body wavefunction \cite{amico08}.
However, quantum correlations and many-body entanglement of a system can appear in different forms among its contituents, and usually is a very hard task to highlight all of its different intricate structures as well as a proper quantification. 

The usual approach deals with the entanglement between two partitions of the system, easily quantified by the von Neumann entropy of the reduced density matrix. 
More recently it was realized that not only the von Neumann entropy of the reduced density matrix has important information about the phase, but also its spectral properties, i.e., the eigenvalues of the reduced density matrix - usually called as entanglement spectrum - can host valuable and more detailed information about the phase \cite{Li2008,Fidkowski2010,Turner2011}. For example, in unconventional topological phases whose entanglement spectrum becomes degenerate due to the presence 
of non abelian edge excitations in the system. Apart from bipartite entanglement, different quantum correlations  quantifiers were also put forward~\cite{amico08,ModiReview2012,HorodeckiReview2009,AlexanderReview2017}, including the analysis of multipartite entanglement \cite{amico08,TRO2006b,Hofmann2014,Konstantin2017,Antonio2020},  pairwise concurrences and quantum discord \cite{ModiReview2012,Beggi2016,Osterloh2002,Iemini2016_concurrence,Campbell2017_concurrence,Lima2021}, particle entanglement \cite{Iemini2015,Alexandradinata2011,Hatem2017,Masudul2009,Herdman2015,Zhao2010,Lukas2017,Masudul2007,Tullio2019}, coherences \cite{AlexanderReview2017,Vianna2016} among others \cite{Chiara2018,bagrov2020}. In a general form, these studies proved very fruitful highlighting how different facets of the correlations shared between the microscopic constituents can be useful and complementary to each other for a complete characterization of quantum systems and their phases of matter. We shall explore one of these facets in this work, specifically, the \textit{particle correlations} shared among its constituents.

In this work, we thus study the one-dimensional extended Hubbard model (EHM) \cite{hubbard63} within the perspective of its \textit{$n$-particle reduced density matrices} ($n$-particle RDM). 
We focus on the $n=2$ case, corresponding to the two-body reduced density matrices. 
We analyse the quantum correlations and many-body entanglement present in the reduced density matrices among the different phases of the model and across its quantum phase transitions. 
Most studies of the EHM in the quantum information context has dealt with the quantum correlations  among the \textit{modes} of the system  \cite{Jian2004,Deng2006,Yang2008,Mund2009,Liu2011,Gigena2013,Barbiero2017,Chung2021,Rausch2020,Pandey2017}. 
Here modes can be any defined set of single particle degrees of freedom, as e.g. the spatial localized degrees along the sites of the chain, or (spatially delocalized) momentum degrees of freedom for single particles.
It was first observed by Shi-Jian Gu et al.\cite{Jian2004} that the entanglement of a single site with the rest of the chain is sensitive to three main symmetry broken phases of the model, 
namely the charge-density-wave (CDW), spin-density wave (SDW) and phase separation (PS).
Further investigation considering the entanglement of spatial blocks with $\ell$ sites and the rest of the chain \cite{Deng2006} showed to be even more sensitive to other phases, as superconducting and bond ordered phases.
In a previous work \cite{Iemini2015}, we started our investigations within this different perspective, i.e., analysing the quantum correlation of the particle reduced density matrix.
Our results focused in the case of $n=1$, showing that the von Neuman entropy of the $1$-particle RDM (usually called as entanglement of particles or fermionic entanglement ~\cite{iemini13b,iemini13a,iemini14,balachandran1,balachandran2,ghirardi02,ghirardi04,schliemann01a,schliemann01b,eckert02,li01,barnum04,somma04,paskauskas01,plastino09,zander10,Debarba2017,Gigena2017,Tullio2018,Gigena2020a,Benatti2020,Benavoli2021}) is  useful for the analysis of the model, capturing its main phase transitions except for subtle transitions between different superconducting forms, and the bond-order wave phase.
Therefore, in this work we take a step beyond the simplest $n=1$ case, considering the more general case with $n=2$, which contains further information
about the correlations and properties of the system~\cite{Gigena2020b}. 
We perform a thorough analysis of the RDM properties using different many-body quantum correlation tools. 
Specifically, we analyse not only its 
(i) von Neumann entropy, quantifying the entanglement of these particles with the rest of the system; 
as well as its (ii) quantum coherence, as a direct manifestation of the quantum superposition principle in the reduced states; 
(iii) entanglement spectrum and entanglement gap, providing a more detailed information of the spectrum structure on the different phases; 
and (iv) its $2$-body cumulant, which is a genuinely two-body correlation matrix, i.e., cannot be described from its $1$-particle RDM.
In a general form, we obtained that these quantifiers are sensitive to most phases of the model, showing peculiar behavior at their quantum phase transitions. 
Depending on the specific phase or quantum phase transitions under scrutinity, the analysis of these quantifiers can be  complementary, providing different facets of the quantum system (e.g. different forms of superconductivity in the model are not easily perceived from quantifiers (i)-(ii)-(iii)-nor from its simpler $1$-particle RDM- while the entanglement spectrum can discriminate it).

\begin{figure}
\centering
\includegraphics[width=\columnwidth]{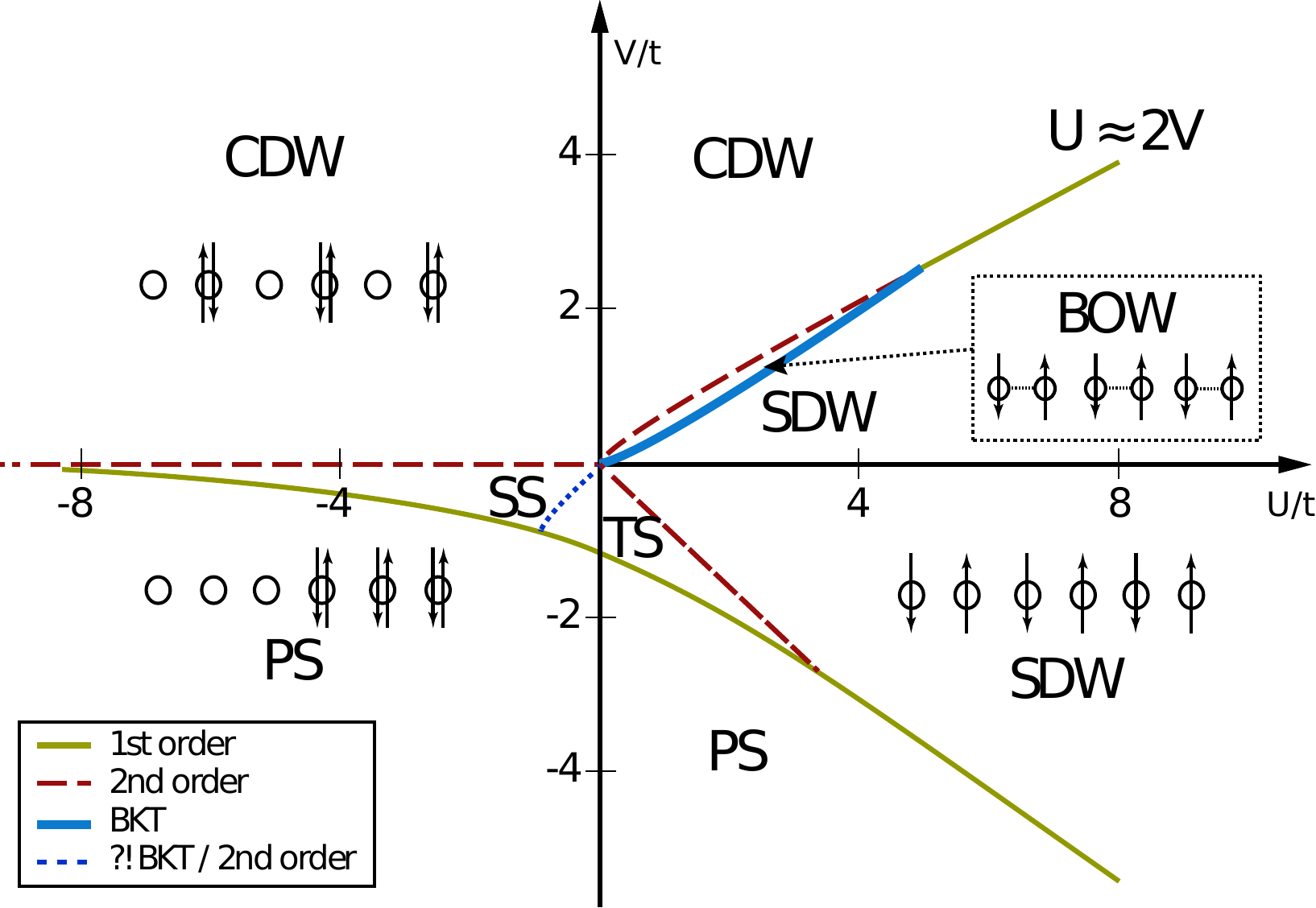}
\caption{ Skematic picture for the phase-diagram of the model, highlighting its different phases and quantum phases transitions as discussed in the literature (see main text). }
\label{phase.diagram.ehm}
\end{figure}

The manuscript is organized as follows. In Sec.~II we introduce the one-dimensional extended Hubbard 
model, its phase diagram at half-filling, and the numerical techniques based on Matrix Product States 
(MPS) used in order to obtain the ground states and correlation functions. 
In Sec.~III we review the definition of $n$-particle reduced density matrices and their properties.
In Sec.~IV we introduce the quantum correlation and entanglement quantifiers studied in this work, 
as well as the concept of entanglement spectrum and entanglement gap.
In Sec.~V we present our results. We first discuss the general qualitative behavior of the quantifiers in the whole phase diagram, and then perform a deeper analysis of finite-size scalings and spectral properties along specific regions in the model. We conclude in Sec.~VI.

\section{Extended Hubbard Model}
\label{sec.ehm} 
 
In this section we review the main properties of the one-dimensional extended Hubbard model.  
All of our studies are focused in the half-filling case. 
The reader familiar with the model might skip to the next section.

The EHM model is a generalisation of the usual Hubbard model \cite{hubbard63, solymon}, 
encompassing broader interactions between the
fermionic particles, such as an inter-site interaction,
thus supporting a richer phase diagram. Precisely, the model is described as,
\begin{eqnarray}
H_{EHM}&=&-t\sum\limits_{j=1}^{L}\sum\limits_{\sigma = \uparrow, \downarrow}
(\ad{j,\sigma}\hat a_{j+1,\sigma} + \ad{j+1,\sigma}a_{j,\sigma}) +\nonumber \\
& & +\, U\, \sum_{j=1}^{L}\hat{n}_{j\uparrow}\hat{n}_{j\downarrow}
 + V\, \sum_{j=1}^{L}\hat{n}_{j}\hat{n}_{j+1},
\label{ehm}
\end{eqnarray}
where $L $ is the lattice size, $\ad{j,\sigma}$ and $\hat a_{j,\sigma}$ are
creation and annihilation operators, respectively, of a fermion with spin
$\sigma = \uparrow, \downarrow$ at  site $j$,
$\hat{n}_{j,\sigma} = \ad{j,\sigma}a_{j,\sigma}$, $\hat{n}_{j}
= \hat{n}_{j,\uparrow} + \hat{n}_{j,\downarrow}$. 
The hopping (tunnelling) between neighbor sites is parametrized by $t$,
while the on-site and inter-site interactions are given by $U$ and $V$, respectively.
We set $t = 1$ as defining our energy scale.
 
Many efforts have been devoted to the investigation of the phase diagram of the EHM at half filling, 
with methods ranging from analytical, pertubative, approximations based on bosonization as well as 
numerical ones~ \cite{lin00,nakamura00,sengupta02,zhang04,dalmonte14,jeckelmann02,Kancharla2001,ejima07,sandvik04,Shinjo2019,Kazuya2020,DaiWei2021}.
Despite the apparent simplicity  of the model it is predicted to exhibit a very rich phase diagram. 
The model can support several distinct phases, namely: spin-density wave (SDW), singlet (SS) and 
triplet (TS) superconductors, phase separation (PS), charge-density wave (CDW) and bond-order wave 
(BOW) - see Fig.\eqref{phase.diagram.ehm} for a sketch of the phase diagram.

In the strong coupling limits, it is intuitive the characterization of its different phases. In the case of strong repulsive onsite interation, $U>0$, $U \gg V$, the fermions avoid double 
occupation and due to the hopping an antiferromagnetic ordering between neighbor sites is formed, 
generating a periodic modulation of spins along the chain, so-called spin density wave (SDW).
The presence of such a phase can be captured by the analysis of the ground state spin correlations 
$\langle \hat \sigma^z_j \hat \sigma^z_\ell\rangle$,
where $\sigma_j^z = \frac{1}{2}(\hat{n}_{j \uparrow}-\hat{n}_{j \downarrow})$.

In the opposite case of a strong repulsive inter-site interaction, $V>0$, $V \gg U$, particles avoid
occupying neighbor sites, tending in this way to occupy the same sites. 
A periodic modulation of charge is now formed, creating  a charge-density wave pattern captured  
by density-density correlations $\langle \hat n_j \hat n_\ell \rangle$ in the ground state wavefunction.

In the case of strong attractive interactions ($U,V<0$ or $U>0$, $V<0$ with $|V|\gg |U|$), 
the particles tend to cluster together and the ground state becomes inhomogeneous with different 
average charge densities in distinct spatial regions. 
Such a phase is called phase separation (PS) and can be observed from the analysis of the charge 
profile along the chain.
 
In the weak coupling limit the analysis becomes subtler, since perturbative arguments might not be 
accurate and intuition might fail.
For small attractive inter-site interactions ($V<0$), superconducting phases are expected to appear, 
characterized by the pairing of fermions which could be observed from pairing correlations. 
The fermions can be paired in different forms, with the possibility of singlet or triplet pairings to occur. 
It is predicted \cite{nakamura00,lin00} a singlet-supercondutor (SS) for approximately $U\leq 2V$, while a triplet-superconducting (TS) for $U \geq 2V$.

The last phase in the model is the controversial bond-order-wave (BOW). 
For small to intermediate values  of positive $U$ and $V$, in a narrow strip between CDW and SDW phases, 
it has been 
predicted~\cite{nakamura00,sengupta02,zhang04,dalmonte14,jeckelmann02,ejima07,sandvik04} 
the appearance of a phase exhibiting alternating  strengths for the expectation value of 
the kinetic energy operator on the bonds, characterized by the order parameter 
$\langle \hat B_{j,j+1} \hat B_{\ell,\ell+1}\rangle$,
where 
$\hat B_{m,m+1} = \sum_{\sigma}(\ad{m,\sigma}a_{m+1,\sigma} + H.c.)$
is the  kinetic energy operator associated with the $m$'th bond.
Such a phase should appear from (i) a continuous CDW-BOW transition; 
and (ii) a Berezinskii-Kosterlitz-Thouless (BKT)   transition from BOW to SDW. 
While from one side the CDW-BOW phase boundary can be  well resolved, 
described by a standard second order phase transition, 
the BOW-SDW boundary is more difficult to locate it precisely, since it involves a BKT transition.
The BKT transition line remains a challenge to delineate and is still subject to debate~\cite{sengupta02,zhang04,ejima07,sandvik04,dalmonte14}. 
 
\textbf{Numerical methods:} In order to obtain numerically the ground states of the EHM and its 
particle reduced density matrices, we use the Matrix Product State (MPS) ansatz, 
which can be a faithful representation of systems in one dimension with local interactions.
This method has established itself as a leading one for the simulation of one-dimensional systems, 
achieving unprecedented precision in the description of static, 
dynamic  and thermodynamic properties for these  systems, 
and quickly becoming the method of choice for numerical studies.
We refer the reader to Ref.~\cite{verstraete2008matrix} for a good review. 
The method can be accomplished by mapping the fermionic model to a spin-half system, i.e., 
representing the fermionic operators with a  Jordan-Wigner transformation~\cite{nielsen2005fermionic} 
that preserves the anti-commutation relations
and thus recovering the usual tensor product Hilbert space structure needed for the implementation of MPS. 
The variational algorithm to minimize the energy was performed using Density Matrix Renormalization Group (DMRG), 
which is standard in such a task. 
In our calculations we used $20$ sweeps in the minimization process, 
which showed enough for an energy convergence of the order of at least $\mathcal{O}(10^{-8})$ and up to 
$\mathcal{O}(10^{-16})$, depending on the region of the phase diagram and the system size.
We also implemented a fast and efficient algorithm to calculate correlators of fourth order, 
needed to construct the $2$-particle RDM's. 
The MPS representation accuracy was  controlled by two parameters, $\chi$ and $D$, 
corresponding to the \textit{minimum allowed singular value permitted} and the \textit{bond link} 
(size of the virtual dimension of the matrices), respectively. 
 Whereas we use an adaptive algorithm which increases the bond link as needed, 
$\chi \approx \mathcal{O}(10^{-20})$ is the minimum singular value considered
and $D=2000$.
All quantities computed in this article have not  significantly changed for larger bond links ($D\sim 4000$), indicating 
a very good precision to the calculations (e.g. the entanglement gap, which is the sutbler quantity under study, has changed its value only at the order of $\mathcal{O}(10^{-12})$ thus validating a good precision).
We consider open boundary conditions in the model because it is best suited to the MPS formalism and finite-size scaling analysis.

\section{n-Particle Reduced Density Matrix}
\label{sec.nparticle.rdm}

In this section we review the definition and some properties of particle reduced density matrices.
In a system of $N$ indistinguishable fermions, described by the set of anticommuting creation 
(and anihilation) operators $\{\ad{i,\sigma}\}$ ($\{\hat a_{i,\sigma}\}$), 
where $i=1,...,L$ stands for the site index and $\sigma = \uparrow,\downarrow$ for the spin index, 
a pure state can always be expanded in the following form,
\begin{equation}
 \ket{\psi} = \sum_{i_{1}\dots i_{N}=1}^{L} 
 \sum_{\sigma_{1}\dots \sigma_{N}=\uparrow,\downarrow} \omega_{i_{1}\sigma_1\dots i_{N}\sigma_1}\ a_{i_{1},\sigma_1}^{\dagger}\dots a_{i_{N},\sigma_N}^{\dagger}\ket{vac},
 \label{staterep}
\end{equation} 
where the coefficients $\omega_{i_{1}\sigma_1\dots i_{N}\sigma_N}$ are antisymmetric in all indices, 
 satisfy the normalization condition of the state and $\ket{vac}$ is the vaccum state.
We can compute the $n$-particle reduced density matrix $\hat \rho_n$ ($1\le n \le N-1$) of $n$  
fermions performing the partial trace over the rest $N-n$ fermions, as follows,
\begin{align}
\hat \rho_{n} = Tr_{(n+1,\dots,N)}\left( \ket{\psi}\bra{\psi}\right).
\end{align}
The partial trace  defines a bipartition $n:N-n$ between $n$ fermions and the rest of the system. 
Usually the calculation of the particle reduced density matrix using 
the partial trace described above can be cumbersome. 
We can, however, obtain it in a different way, which will turn useful for our purposes. 
Instead of taking the partial trace one can compute all $n$-body correlators, 
which correspond to the matrix elements of $\hat \rho_{n}$, 
and in this way reconstruct the reduced state as in a tomographic process. 
In other words, the reduced density matrices of one and two fermions have the following entries:
\begin{align}
\left[\hat \rho_{1}\right]_{\left(i\sigma_{i}\right),\left(j\sigma_{j}\right)} &= \binom{N}{1}^{-1}\bra{\psi} a_{i\sigma_{i}}^{\dagger}a_{j\sigma_{j}}\ket{\psi},\\
\left[\hat \rho_{2}\right]_{\left(i\sigma_{i}j\sigma_{j}\right),\left(k\sigma_{k}\ell\sigma_{\ell}\right)} &= \binom{N}{2}^{-1}\bra{\psi} a_{i\sigma_{i}}^{\dagger}a_{j\sigma_{j}}^{\dagger}a_{k\sigma_{k}}a_{\ell\sigma_{\ell}}\ket{\psi},\label{eq.2body.rdm.definition}
\end{align}
where $\binom{N}{M}$ is the binomial coefficient. 
One can  also obtain the $1$-particle reduced density matrix from an 
integration of the $2$-particle reduced density matrix, 
\begin{equation}
 \left[\hat \rho_{1}\right]_{\left(i\sigma_{i}\right),\left(j\sigma_{j}\right)} = \mathcal{N}
 \sum_{k,\sigma_k}
 \left[\hat \rho_{2}\right]_{\left(i\sigma_{i} k\sigma_{k}\right),\left(k\sigma_{k} j\sigma_{j}\right)} 
\end{equation}
with $\mathcal{N}=1/2$ the normalization constant.
In general, an $n$-particle RDM can always be obtained from its higher orders $(k>n)$-particle RDM 
from a proper integration over its tensor elements. The inverse is obviously not true. 
It is important to recall, however, that the elements of higher orders $(k>n)$-particle RDM are 
partially related to the elements of their lower orders, apart from their cumulants \cite{kubo62}. 
Specifically, for the case of $2$-particle RDM, we have that its elements 
can be expanded in the following form,
\begin{eqnarray}
 \binom{N}{2}\left[\hat \rho_{2}\right]_{\left(i\sigma_{i}j\sigma_{j}\right),\left(k\sigma_{k}\ell\sigma_{\ell}\right)}
 =
 N^{2}\left[\hat \rho_{1}\right]_{\left(i\sigma_{i}\right),\left(j\sigma_{j}\right)}
 \left[\hat \rho_{1}\right]_{\left(k\sigma_{k}\right),\left(\ell\sigma_{\ell}\right)}
 - \nonumber \\
 N^{2}\left[\hat \rho_{1}\right]_{\left(i\sigma_{i}\right),\left(\ell\sigma_{\ell}\right)}
 \left[\hat \rho_{1}\right]_{\left(k\sigma_{k}\right),\left(j\sigma_{j}\right)}
  + \left[\hat \Delta_2 \right]_{\left(i\sigma_{i}j\sigma_{j}\right),\left(k\sigma_{k}\ell\sigma_{\ell}\right)} 
  \label{eq.cumulant}
\end{eqnarray}
where $\hat \Delta_2$ is the $2$'nd order cumulant, 
corresponding to the elements of $\hat \rho_{2}$ that cannot be obtained from lower orders $\rho_{n<2}$. 
We can see that cumulants are Hermitian matrices.

\section{Quantum correlations, Entanglement and Coherence}
\label{sec.qc}

\begin{figure*}
\centering
\includegraphics[scale=1.0,width=0.95 \columnwidth]{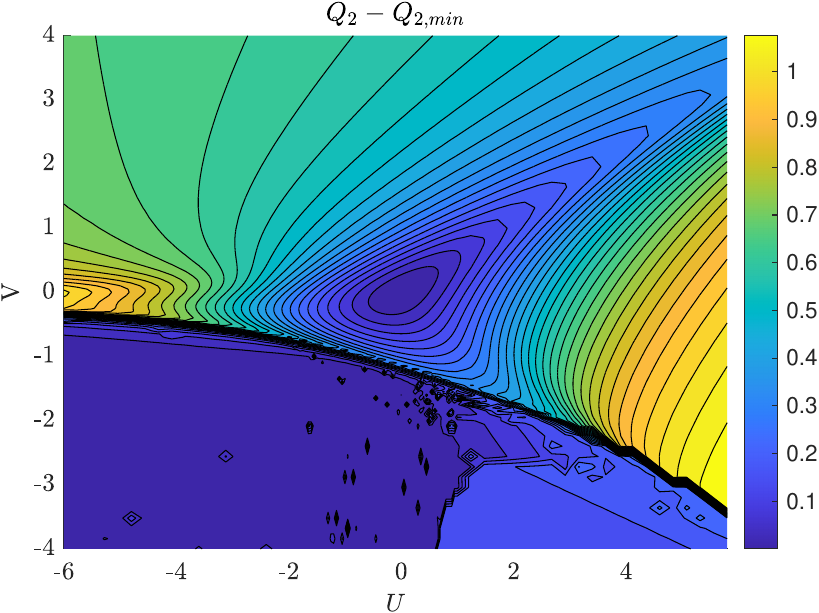}
\includegraphics[scale=1.0,width=0.95\columnwidth]{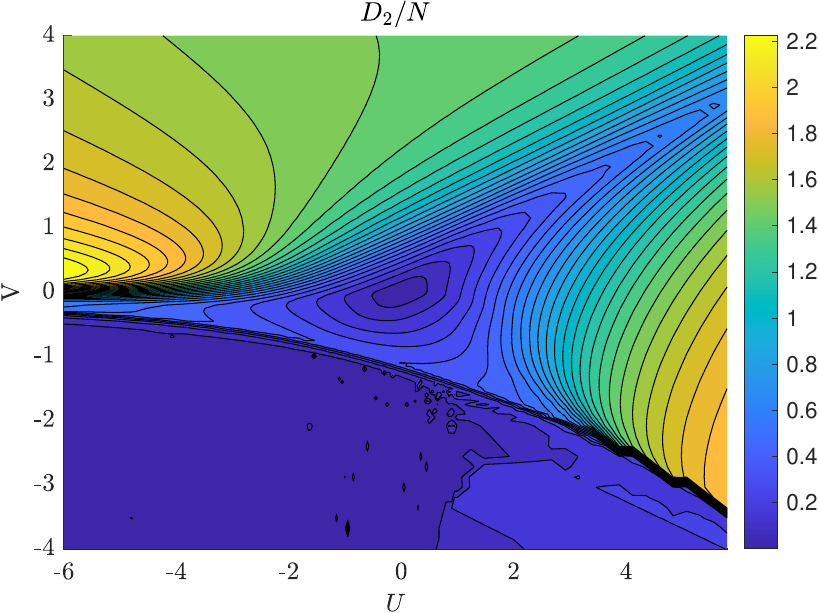}
\includegraphics[scale=1.0,width=0.95\columnwidth]{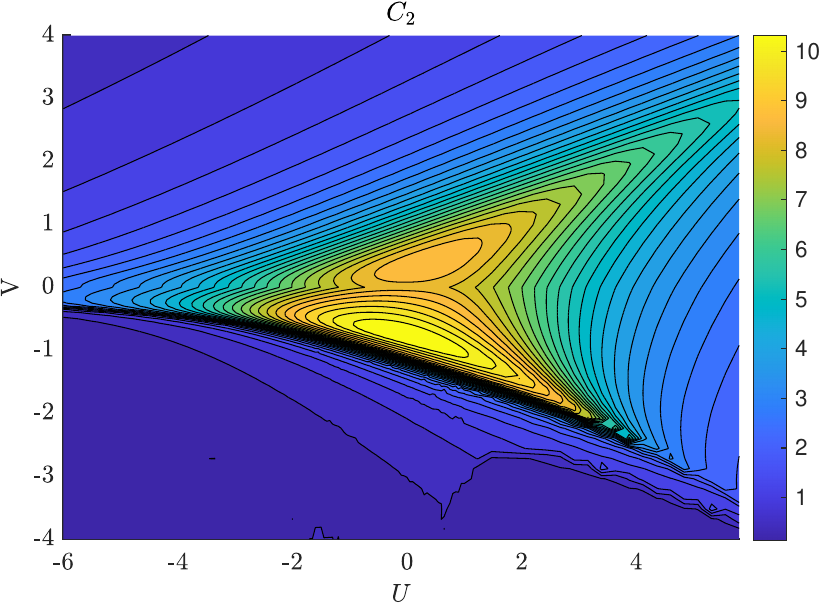}
\includegraphics[scale=1.0,width=0.95\columnwidth]{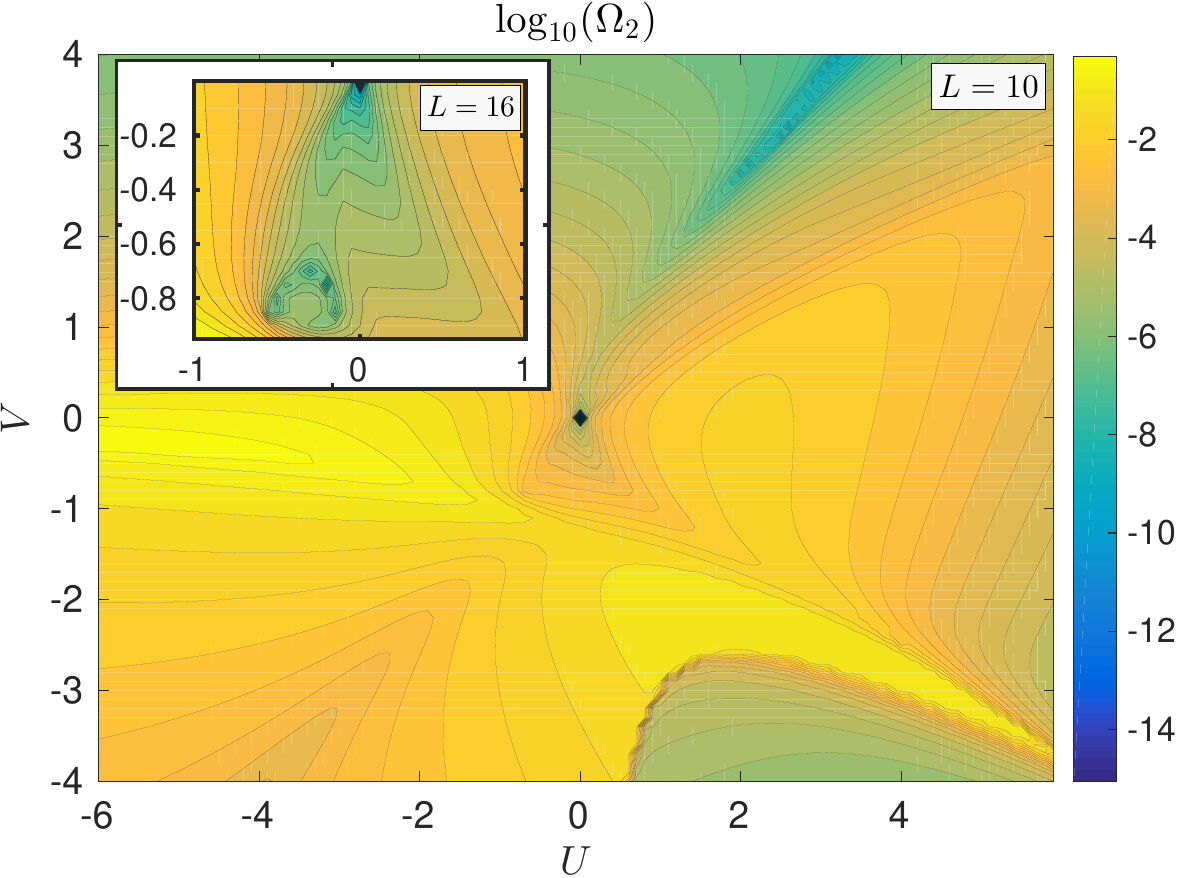}
\caption{ Results for the two-body reduced density matrix in a system at half filling ($N=L$ fermions) along the full phase diagram of the model. We show in \textbf{(top-left panel)} the quantum correlations $Q_2$, \textbf{(top-right panel)} irredutible two-body correlations $D_2$,
\textbf{(bottom-left panel)} coherence $C_2$, and 
\textbf{(bottom-right panel)} the entanglement gap $\Omega_2$. 
Except for the bottom-right panel (entanglement gap $\Omega_2$), we consider a system with $L=16$ sites. In the entanglement gap panel we consider a system with $L=10$ sites, and highlight in the inset its behavior for larger $L=16$ sites across the supercondutcting region.
 We see that all these quantities capture most of the quantum phase transitions of the model, e.g. displaying discontinuities at $1$st order transitions and continuous maximum/minimum values at $2$nd order transitions.
}
\label{phase.diagram.q2}
\end{figure*}

In this section we review the definition and properties of the quantum correlations and 
entanglement quantifiers studied in the paper.

\textbf{Quantum correlations:} Perhaps the most familiar quantum information concept 
which has proved a powerful tool in the study of quantum correlations in many-body system 
is the well known von Neumann entropy.
Given a pure state, the von Neumann entropy of a reduced density matrix has information 
about the quantum correlations between the partition and the rest of the system. 
In systems of indistinguishable particles, 
a partition of the system could be defined in different forms: 
(i) a partition between two sets A and B of modes of the system, 
performed through partial trace over one of the sets, or 
(ii) a partition between the $n$ and $N-n$ particles of the system, 
performed through the partial trace of $N-n$ particles on the state.
The first approach provides information about the quantum correlations between the modes of the system, 
while the second one concerns the quantum correlations among the particles
\cite{iemini13b,iemini13a,iemini14,balachandran1,balachandran2,ghirardi02,ghirardi04,schliemann01a,schliemann01b,eckert02,li01,barnum04,somma04,paskauskas01,plastino09,zander10,Debarba2017,ding2020,ding2021}.
These two notions of quantum correlations are complementary, 
and the use of one or the other depends on the particular situation under scrutiny. 
For example, if one is interested in certain quantum information protocols a description in terms of 
modes might be more appropriate, 
while correlations in eigenstates of a many-body Hamiltonian could be more naturally described by its 
particle perspective.
In this work we deal exclusively with the particle framework.

We define in this way the quantum correlations between the set of $n$ and $N-n$ particles 
in a pure state as,
\begin{equation}
 Q_n \left (\ketbra{\psi}{\psi} \right) = 
 S(\hat \rho_n),
\end{equation}
where $S(\hat \rho_n ) = - Tr(\hat \rho_n \log(\hat \rho_n))$ is the von Neumann entropy 
of the $n$-particle reduced density matrix. It is worth making a few observations.
Since the system is composed of indistinguishable fermions, 
due to the antisymmetrization of the wavefunction, $Q_n$ is never null. 
However, for states described by a single Slater determinant the quantum correlations have 
a minimum given by $Q_{n,\min} =  \ln  \binom{N}{n}$, 
while $Q_n $ is larger for any state which cannot be described by a single Slater determinant. 
It leads us to the conclusion that the minimum $Q_{n,\min}$ corresponds simply to the 
exchange correlations due to the antisymmetrization postulate, 
the difference $Q_n - Q_{n,\min}$ being the significant term.
 
One can show \cite{plastino2009,plastino2016} that the von Neumann entropy of the reduced state, 
and consequently our quantum correlation quantifier, is bounded as follows:
\begin{equation}
 \ln  \binom{N}{n} \leq Q_n \left (\ketbra{\psi}{\psi} \right) \leq \ln \binom{d}{n}
\end{equation}
where $d$ is the number of single-particle degrees of freedom in the system ($d=2L$ in our system). 
The minimum is reached if and only if the pure state $|\psi\rangle$ can be described by a single Slater determinant.
 
\textbf{Quantum coherence:} In quantum mechanics the coherence of a state is a direct 
manifestation of the quantum superposition principle. 
Despite its fundamental importance in quantum theory, 
only more recently its proper quantification and characterization have been formalized 
\cite{Baumgratz2014}, and a few different measures of quantum coherence were proposed. 
In this work we concentrate on the analysis of the \textit{$\ell_1$ norm of coherence}, 
defined by the integration of the absolute value of the off-diagonal matrix elements,
\begin{equation}
 C_{\ell_1}^{[n]}(\hat \rho_n) =  \sum_{i \neq j} |(\hat \rho_n)_{i,j}|.
\end{equation}
We notice that coherence quantifiers are basis dependent.

\textbf{Entanglement spectrum:} The entropy of the reduced density matrix, as discussed previously, 
provides useful information about the correlation among the contituents of the system. 
It was realized however that further insights about the many-body properties of the system can be 
obtained from its spectral structures.
Specifically, given the reduced density matrix $\hat \rho_n$, 
it can be diagonalized as 
$\hat \rho_n = \sum_i e^{-\xi_i} \ketbra{i}{i}$, with $\xi_i \leq \xi_{i+1}$. 
Writing $\hat \rho_n = e^{-\hat H_n}$, 
we see that $\xi_i$ and $\ketbra{i}{i}$ can be regarded as eigenvalues and eigenvectors 
of a fictitious $n$-body Hamiltonian $\hat H_n$. 
The entanglement spectrum $\{\xi_i\}$ is in this way interpreted as the eigenvalues 
associated to the parent Hamiltonian $\hat H_n$. 
Many efforts have been devoted studying the entanglement spectrum in spatial partitions of ground state wavefunctions, 
leading e.g. to a better understanding of bulk-edge properties in topological insulators and superconductors 
\cite{Li2008,Fidkowski2010,Turner2011}.

We study in this article the entanglement spectrum in particle partitions, 
a subject much less explored so far \cite{Zhao2010,Herdman2015,Masudul2009} 
(see also \cite{Rex2014} for momentum partitions or 
\cite{Alexandradinata2011,Sterdyniak2011} for hybrid spatial/particle partition approaches),
focusing on its entanglement gap,
\begin{equation}\label{eq:ent.gap}
 \Omega_n = \xi_2 - \xi_1.
\end{equation} 
and analysis of the dominant eigenvector of the reduced state, i.e., the one corresponding to the largest eigenvalue.

\textbf{Cumulant matrix:} The cumulant matrix $\hat \Delta_2$, as defined in Eq.\eqref{eq.cumulant}, 
contains the two-particle information that cannot be obtained 
from the single particle reduced density matrix ($\hat \rho_1$), 
or in other words, from single-particle observables. 
It describes in this way fundamental two-particle correlations in $\hat \rho_2$, 
also called as irreducible two-particle correlations \cite{Werner1999}.
We study the contribution of such correlations from the $\ell_1$ norm of the cumulant matrix, 
defining the $n$-particle irreducible correlation as follows,
\begin{equation}
 D_{n} \equiv ||(\hat \Delta_n)||_{1} 
 = \sum_i |\lambda_i|
\end{equation}
with $\lambda_i$ the eigenvalues of the cumulant matrix.
It is worth noting that for states described by a single Slater determinant, 
cumulants of any order vanish \cite{Werner1999}, $\hat \Delta_{n} = 0$ for $2\leq n \leq N-1$, 
thus leading to null particle irreducible correlations.

\section{Results}
\label{sec.results}

In this section we present our results for the model. In a general way, we obtained that our quantifiers  capture 
most of the quantum phase transitions of the model, see Fig.\eqref{phase.diagram.q2}), with exception to some BOW related phase transitions.
The quantifiers show peculiar behaviors such as discontinuities, maximum or minimum values at the quantum phase transitions. 
It is interesting to interpret such behaviors based on the order parameters for the different phases of the model. As discussed in Sec.\eqref{sec.ehm}, the different phases are characterized by their order parameters $\hat O_i$ (e.g. charge-operator (CDW), spin-operator (SDW) and others) and corresponding correlators $\langle \hat O_i \hat O_j \rangle$. These correlators correspond to specific elements of the RDM.
On the verge of a second-order phase transition the correlator's correlation length, which is an implicit function of the Hamiltonian gap, tends to diverge. Therefore, these terms will be dominant in the RDM and a peculiar behavior of our quantifiers is also expected along these transitions, corroborating with our numerical results. 
Although there is an implicit connection between the order parameter correlators, the Hamiltonian gap and our quantifiers, the latter will in general correspond to intricate functions of the RDM elements, thus does not necessarily implying in a clear (linear) connection among all these properties. 
In fact we find (as we discuss in more detail below) that some quantifiers may be more sensitive to certain transitions than others, thus working in a complementary form in order to  describe the phase diagram of the model.

In Fig.\eqref{phase.diagram.q2}-(top panels) we show the quantum correlations ($Q_2$) and 
irredutible correlations ($D_2$) for the phase diagram of the model. 
We see that both the quantum correlations of $2$ fermions with the rest of the $N-2$ particles ($Q_2$) as well the correlation between the reduced $2$ fermions ($D_2$) in the reduced state behave qualitatively similar, showing discontinuities at the $1$'st order transitions of the model, while are continuous reaching minimum values at the $2$'nd order phase transitions. 

In Fig.~\eqref{phase.diagram.q2}-(bottom left) we show the 
quantum coherence ($C_2$) in the reduced density matrices. We recall that the coherence here is computed in the real space basis.
While at $1$st order transitions it also display discontinuities, 
at the $2$nd order phase transitions it presents maximum values. 
Since coherence is a basis dependent quantity, 
and we work at the real space representation for the reduced density matrix, 
at the quantum phase transitions we expect it to be maximum due to the divergence of the coherence length.  

We show in Fig.~\eqref{phase.diagram.q2}-(bottom right) the entanglement gap ($\Omega_2$).
Even though the eigenvalues of the reduced density matrix can be gapless, the two ``dominant'' excitations contain relevant information of the phase, 
in the same spirit as  Penrose-Onsager crtierion \cite{Shi2003,Patrycja2019}. 
We see that the entanglement gap 
display a similar behavior of maximum/minimum and discontinuities along the transitions of the model.  
We further notice a peculiar behavior of the entanglement gap within the superconducting phase (see inset panel of Fig.~\eqref{phase.diagram.q2}), suggesting a phase transition, or a change of dominant eigenvalue with the gap closing/crossing. We devote a more detailed analysis of this point 
in Sec.~\ref{subsec.superconducting}, highlighting the presence of a TS/SS superconduting transition. Moreover, the entanglement gap also displays an anomalous behavior in the strong coupling regions $U/V \sim 1$ with $t \ll 1$ , which is not seen in the other quantifiers nor expected from the known phase diagram of the model (Fig.\eqref{phase.diagram.ehm}). While for $U,V>0$ we see a very abrupt closure of the entanglement gap, in the opposite case with $U,V<0$ there is a less apparent (but still emergent) minimum in the quantifier. 
We attribute these behaviors  to the different ``defects'' (at a few-body level) that can occur in the corresponding phases, as we discuss  in  Sec.\eqref{subsec.u.sim.v}. Depending on the ratio $V/U$ different types of local defects prevail in the ground state wavefunction. 
The entanglement gap, interestingly, is more sensible to such few-body fluctuations in the wavefunction as compared to the other quantifiers. This increased sensitivity may be a  consequence of its definition, based on a restricted (the dominant) set of eigenvalues/eigenvectors of the RDM. In this way it can work as a magnifying glass on specific changes over the RDM such as those caused by few-body fluctuations, differently from the other quantifiers which are complex functions integrated over all degrees of freedom of the reduced density matrix. Since fluctuations at a few-body level are suppressed over the full degrees of freedom, they become less apparent for such quantifiers.

We discuss now in more detail the behavior of the quantifiers along specific regions of interest in the phase diagram.

\subsection{$U/t = 4$}

\begin{figure*}
\includegraphics[scale=1.0,width=\columnwidth]{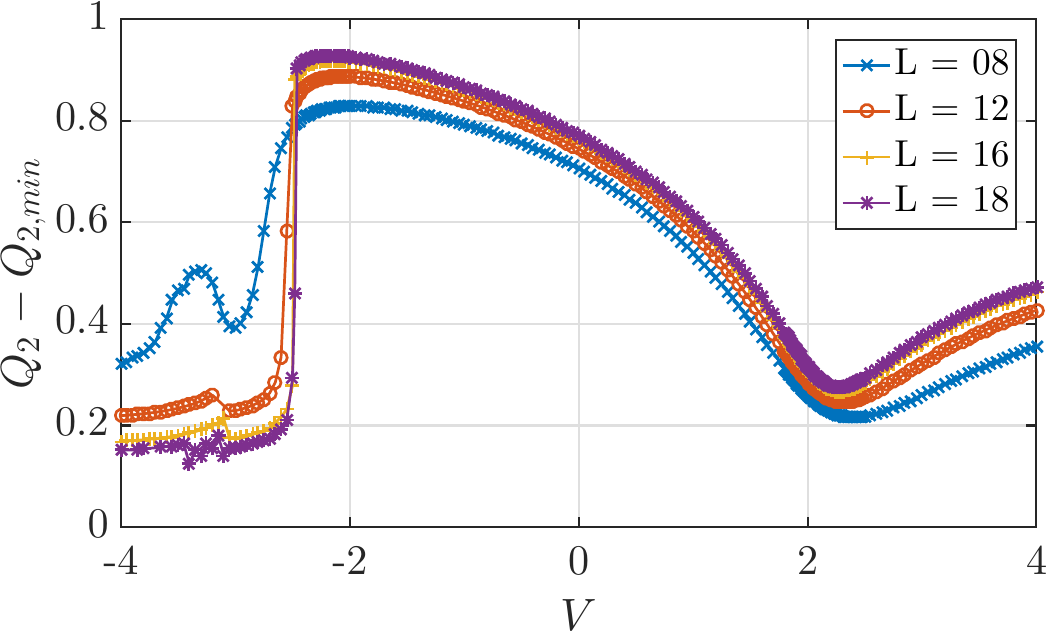}
\includegraphics[scale=1.0,width=\columnwidth]{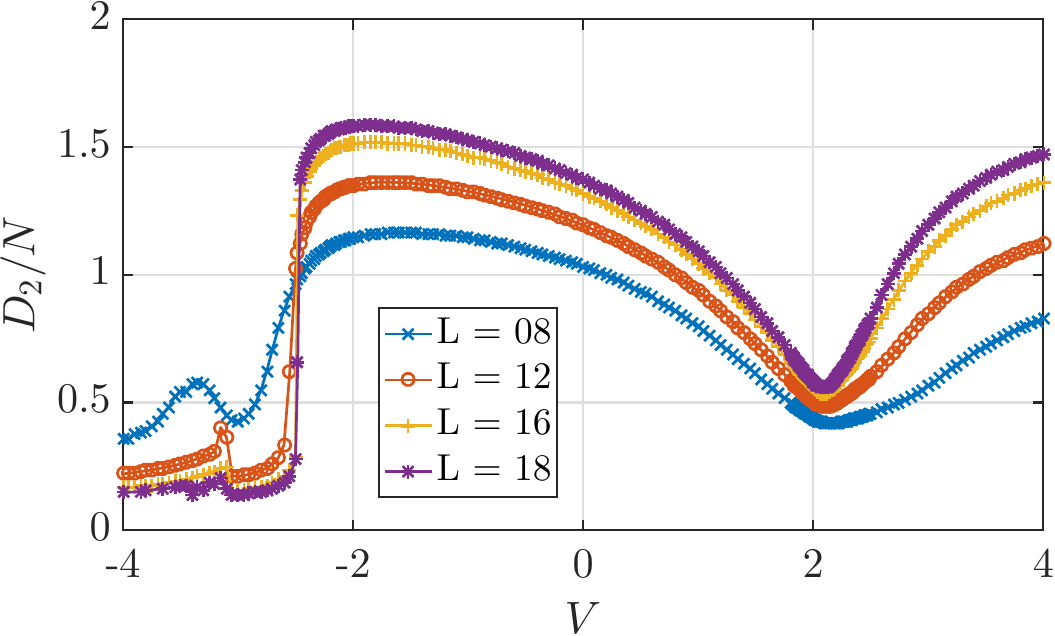}
\includegraphics[scale=1.0,width=\columnwidth]{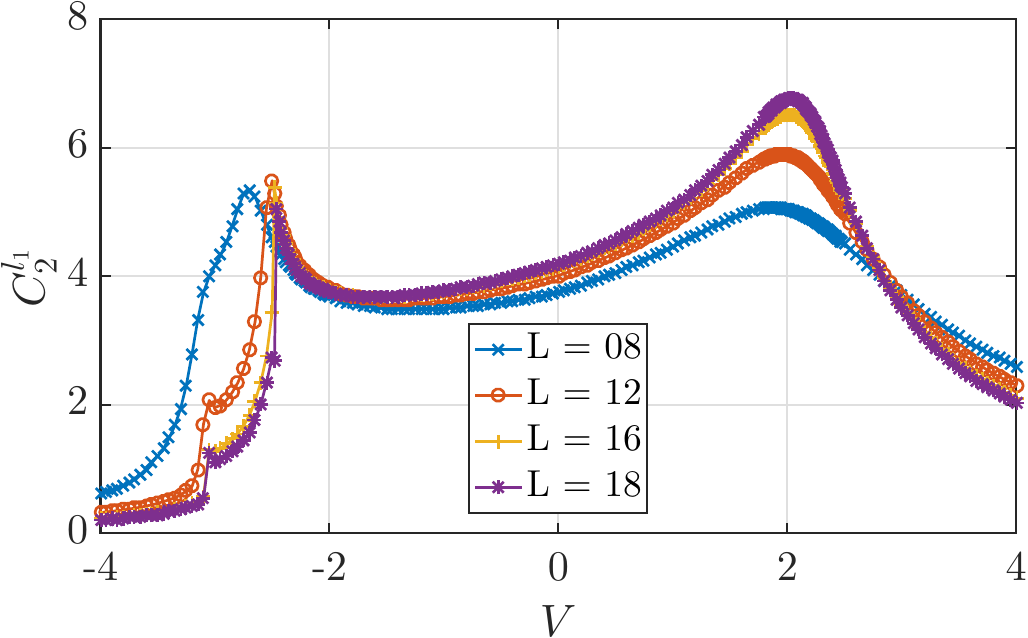}
\includegraphics[scale=1.0,width=\columnwidth]{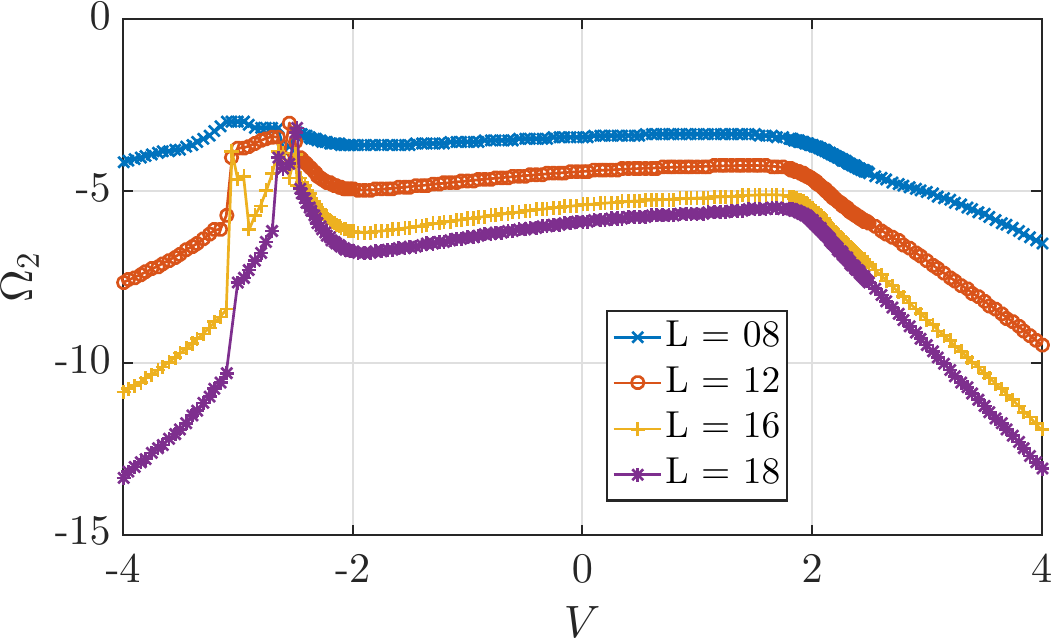}
\includegraphics[scale=1.0,width=\columnwidth]{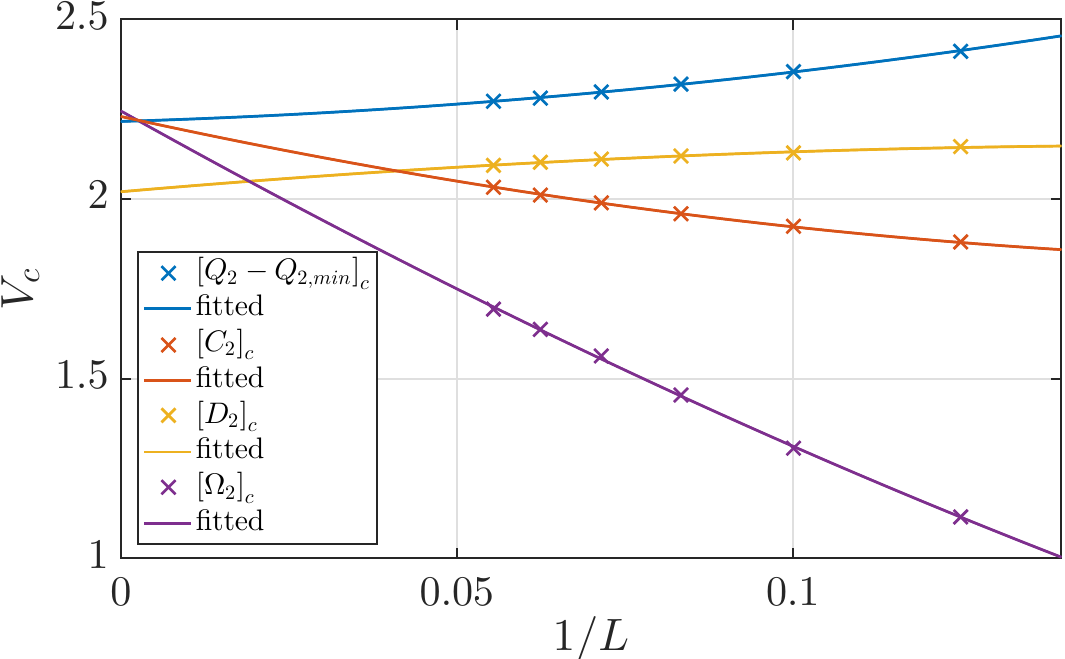}
\caption{
Results for the two-body reduced density quantifiers along the line in the phase diagram with $U/t=4$ fixed, for varying $V/t$ and system sizes $L$, highlight transitions between PS1-PS2-SDW-CDW. 
We show in \textbf{(top-left panel)} the quantum correlations $Q_2$, \textbf{(top-right panel)} irredutible correlations $D_2$, 
\textbf{(middle-left panel)} coherence $C_2$ and  
\textbf{(middle-right panel)} entanglement gap $\Omega_2$. In the 
\textbf{(bottom panel)} we show the finite-size scaling analysis for the interactions $V(L)^*_{[...]}$ where the quantifiers are maximum, or minimums, in the region with $U/t > 0$. The fitted lines use a second order polynomial in $1/L$.
}
\label{fig.linhaU4.varV}
\end{figure*}

Along the line with fixed $U/t = 4$ and varying inter-site interactions $V/t$, the model shows differente phase transitions, namely, PS-SDW, SDW-BOW and  BOW-CDW transitions. We show in Fig.\eqref{fig.linhaU4.varV} our quantifiers along this line, for different system sizes. We see the behaviors for the quantifiers discussed previously. A few aspects are worth remarking. The quantifiers show two discontinuities along the PS phase. This is due to the existence of different PS phases in this region, where the fermions tend to cluster in two different structures, as discussed also in Refs.\cite{lin00,Deng2006}.

We perform a finite-size scaling analysis for the interaction $V^*(L)_{[...]}$ where the quantifiers are maximum/minimum along the line, in the region $U/t>0$, and compare with expected results for the critical interaction of the literature. We show our results in Fig.\eqref{fig.linhaU4.varV}-(bottom panel). We obtain that,
\begin{eqnarray}
 V^*(L \rightarrow \infty)_{[Q_2]} \cong 2.22,&\,& 
 V^*(L \rightarrow \infty)_{[C_2]} \cong  2.23, \nonumber \\
 V^*(L \rightarrow \infty)_{[D_2]} \cong  2.02,&\,& 
 V^*(L \rightarrow \infty)_{[\Omega_2]} \cong  2.24,\nonumber \\
 & &
\end{eqnarray}

It is interesting put these results in perspective with those obtained in the literature. According to the literature, the best estimates for the quantum phase transitions in this region correspond to $V/t \approx 2.16$~\cite{sengupta02,zhang04,jeckelmann02,ejima07,sandvik04}   for the CDW-BOW transition, and  $V/t \approx 1.88-2.00$~\cite{sengupta02,zhang04,ejima07,sandvik04} or  $V/t \approx 2.08$~\cite{dalmonte14} for the BOW-SDW transition. We see in this that while $Q_2, C_2$ and $\Omega_2$ are close to the expected CDW-BOW transition point, $D_2$ is closer to the BOW-SDW transition.

\subsection{Superconducting phase}
\label{subsec.superconducting}

We focus here in the analysis of the superconducting phase of the model. It is convenient to first discuss a few symmetries of the model and the two-body reduced density matrix. We first define the total spin operator $\vec{S}^2$ and total spin along $z$-axis, respectively, as
\begin{eqnarray}
  \vec{S}^{2} &=& \frac{1}{2}\hat N+\frac{1}{4}\sum_{ij}^{L}\left(\hat n_{i \uparrow}- \hat n_{\downarrow}\right) \left(\hat n_{j\uparrow }- \hat n_{j\downarrow }\right) + \nonumber \\
  &-&\sum_{ij}\ad{i \uparrow}\ad{j\downarrow}a_{i\downarrow}a_{j \uparrow} \\
  \hat S_z &=& \frac{1}{2}\sum_i (\hat n_{i\uparrow }- \hat n_{i\downarrow })
 \end{eqnarray}
 with $N$ being the total number operator. It is not hard to see that the Hamiltonian commutes with the above operators, thus possesing a $su(2)$ symmetry \cite{essler2005}. It is not direct that the two-body reduced density matrix should inherit the symmetries of the Hamiltonian. We notice, however, from its own definition that terms that do not conserve the total spin along $z$ direction are null. Thus the reduced density matrix inherits at least the symmetry $\hat S_z$.
 
\begin{figure}
\includegraphics[scale=1.0,width=\columnwidth]{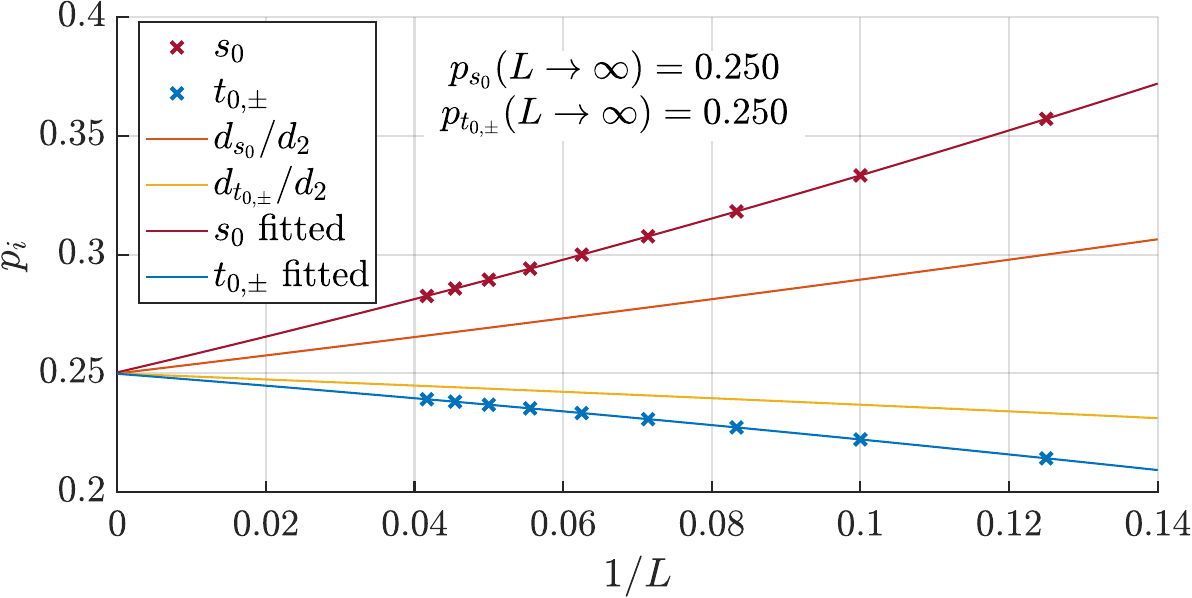}
\caption{Asymptotic limit of the subspaces probabilities as a function of $1/L$. The fitted lines use a second order polymial in $1/L$.}
\label{fig.prob}
\end{figure}

 Interestingly, we observed numerically that the two-body reduced density has also symmetry $\vec{S}^2$, therefore indeed sharing the $su(2)$ Hamiltonian symmetry. We were not able, however, to demonstrate it analytically, rather we observed numerically along all phase diagram and for different system sizes that this property is present. 
 The $su(2)$ symmetry in the reduced state leads to interesting consequences and avenues of investigation for the analysis of the state in the superconducting phase.
 We first recall that the total spin operator commutes with $\hat S_z$ and split the Hilbert space of the reduced density matrix into triplet and singlet subspaces with quantum number $(S^2,S_z)$ given by,
 \begin{eqnarray}
  t_\pm \equiv (2,\pm 1), \quad t_0 \equiv (2,0),\quad s_0 \equiv (0,0)
 \end{eqnarray}
where $t (s)$ denotes triplet (singlet) subspace. The dimension of the antisymmetric subspace of the  Hilbert space for two particles corresponds to $d_2 = \binom{2N}{2}$. The fraction of the dimension for the singlet on such space is given by,
\begin{equation}
  \frac{d_{s_0}}{d_2}= \frac{N(N+1)}{2}\binom{2N}{2}^{-1} 
  = \frac{N+1}{4N-2}
 \end{equation}
 which in the thermodynamic limit $N\rightarrow\infty$ reduces to $
 \lim_{N\to\infty}  d_{s_0} /d_2 = 1/4$. 
Similarly, for triplet subspaces $t_0,t_\pm$ we have,
\begin{eqnarray}
  \frac{ d_{t_0,t_\pm} }{d_2} =\frac{N-1}{4N-2} \overset{N\rightarrow \infty }{\longrightarrow } 1/4
 \end{eqnarray}
 All subspaces converge to the same ratio of $1/4$ in the thermodynamic limit. The symmetries imply in a block diagonal structure for the reduced state in the above subspaces, and can be written as,
 \begin{equation}
  \hat \rho_2 = \sum_{i = s_0,t_0,t_{\pm}} p_i ( \mathcal{P}_i \hat \rho_2 \mathcal{P}_i^\dagger)
 \end{equation}
 with $\mathcal{P}_i$ the projectors onto the singlet and triplets subspaces, and $p_i = Tr(\hat \rho_2 \mathcal{P}_i)$ the overlap of the reduced density matrix in its respective subspace.
We obtain numerically that the overlap $p_i$ of the reduced density matrix on each subspace is constant along all phase diagram of the model, depending only on the number of sites in the system. Moreover, in the thermodynamic limit the overlap of all subspaces tend to the fraction of their dimensions over the antisymmetric Hilbert space, precisely, $p_i \rightarrow d_{i}/d_2=1/4$ for 
$N \rightarrow \infty$ with $i = s_0,t_0,t_{\pm}$ - see Fig.\ref{fig.prob}. 

\begin{figure}
\includegraphics[scale=1.0,width=\columnwidth]{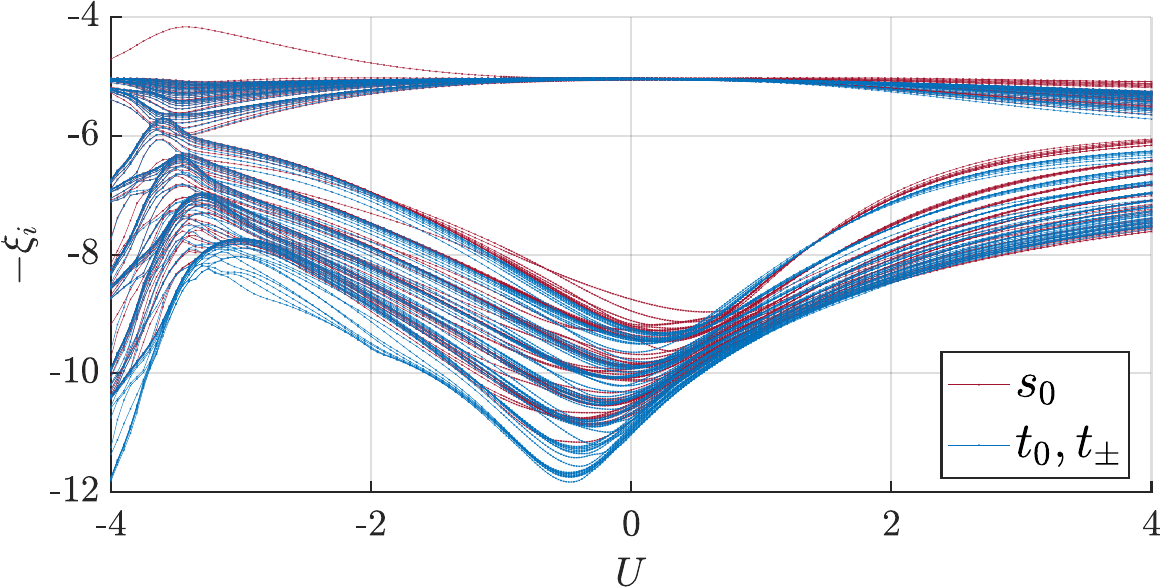}
\includegraphics[scale=1.0,width=\columnwidth]{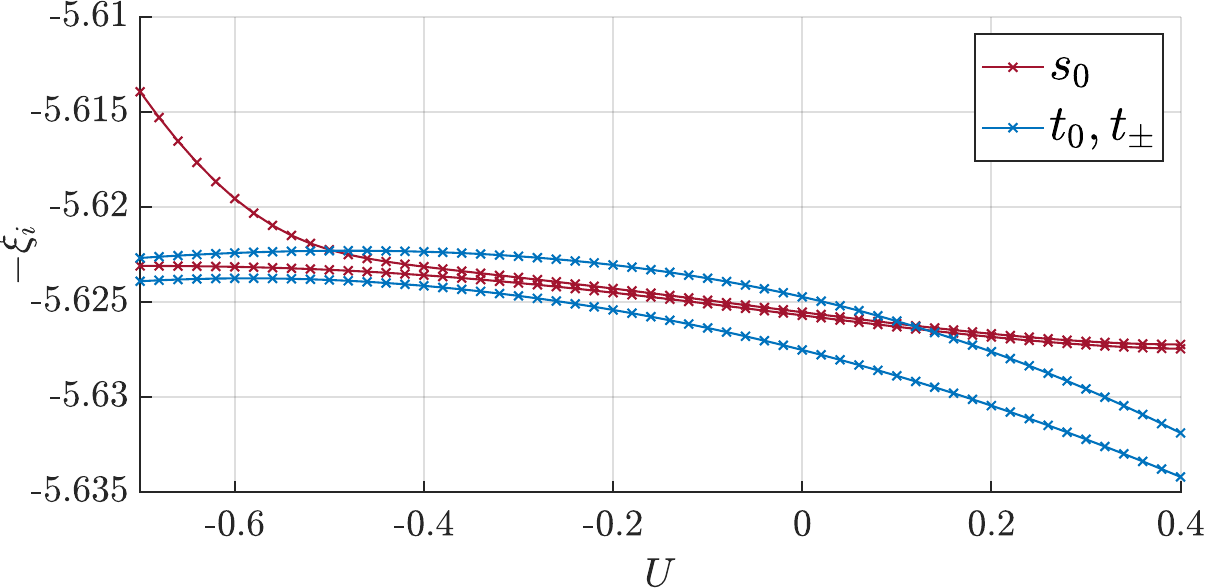}
\caption{ Entanglement spectrum of the reduced density matrix $\hat \rho_2$ for a system with $L=18$ ($L=24$) sites and interactions $V/t=-0.5$ ($V/t=-0.6$) for the top (bottom) panel. We show here only the first $100$ ($8$) largest eigenvalues of the spectrum.}
\label{fig.spectrum.superconducting}
\end{figure}

\textit{Spectral properties.} We study the spectral properties of the reduced density matrix, taking in consideration the spliting of  singlet and triplet quantum numbers. 
We show in Fig.\eqref{fig.spectrum.superconducting} the largest eigenvalues of the reduced density matrix in the superconducting region, for varying system sizes and on-site interactions. 
We see that around  $U/t \sim -3$, with $V/t = -0.5$,  there is a single dominant eigenvalue, corresponding to the singlet subspace. Indeed in this region we expect, according to the literature, the existence of a SS phase. As the on-site interaction is decreased (in modulus), triplet eigenvalues become comparable to the dominant singlet, until at a certain interacting value ($U^*_{\rm{ss-ts}}(L,V/t)$) the triplet dominant eigenvalue surpass the singlet and becomes the largest eigenvalue. In this region we expect the dominance of a TS phase. The existence  of different superconducting orderings was also studied in Refs.\cite{nakamura00,lin00,Shinjo2019,Kazuya2020}.

\begin{figure}
\includegraphics[scale=1.0,width=0.95\columnwidth]{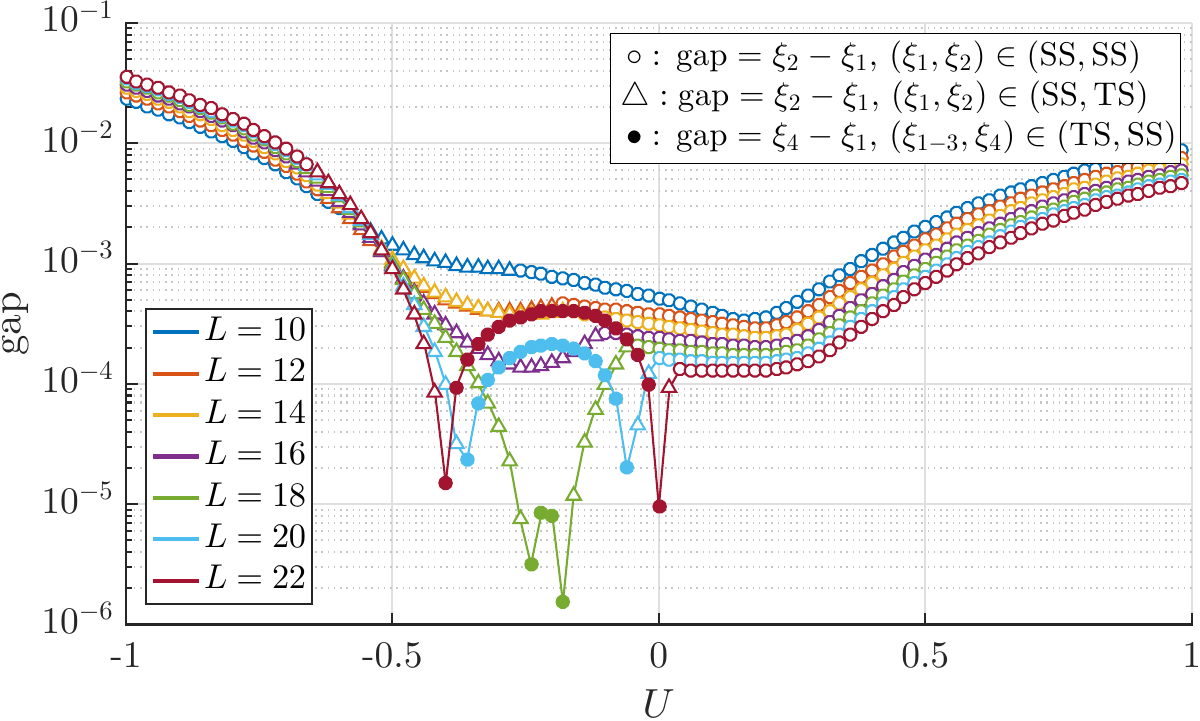}
\includegraphics[scale=1.0,width=0.95 \columnwidth]{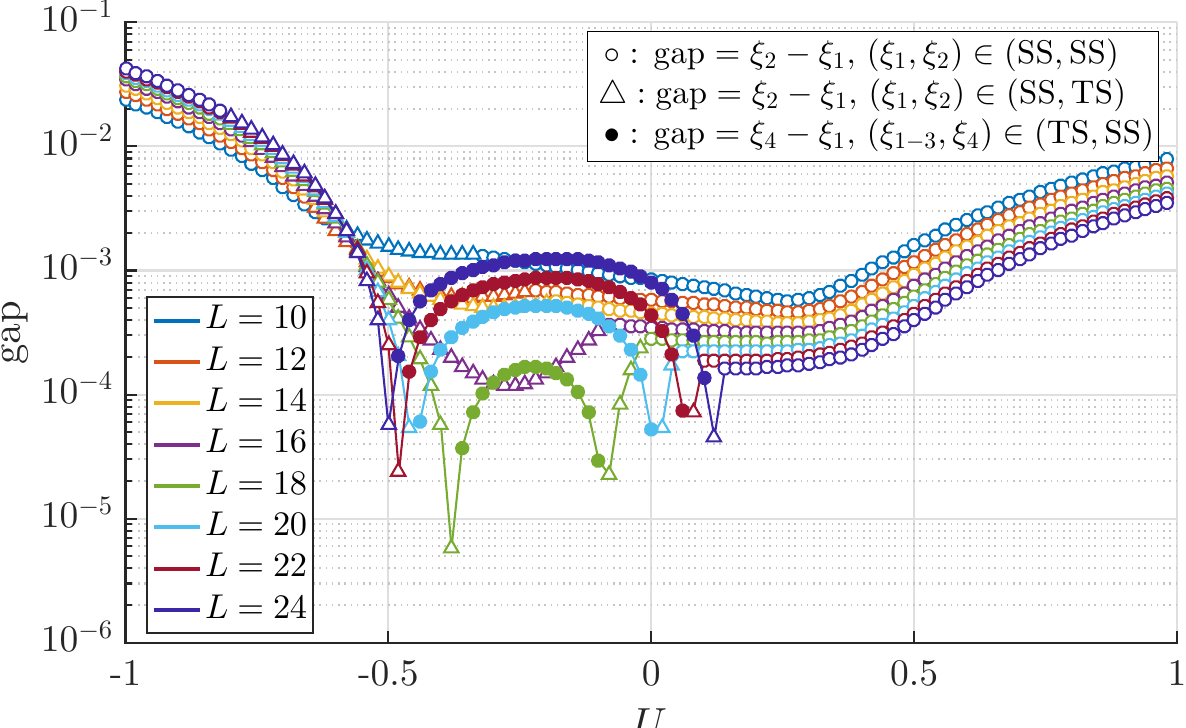}
\caption{ Entanglement spectrum gap for the lines $V/t=-0.5$ (top) and $V/t=-0.6$ (bottom), for varying system sizes and on-site interactions. 
Empty (filled) symbols denotes a dominant singlet (triplet) eigenvalue. In the case where the triplet is the dominant eigenvalue  we have a three-fold degeneracy in the largest eigenvalues ($\xi_{1} = \xi_{2,3}$) all belonging to triplet subspaces, we show in this case the gap with the next largest eigenvalue ($\xi_4$), according to the figure legend.}
\label{fig.line3.gap}
\end{figure}

In Fig.\eqref{fig.line3.gap} we show the gap $\Omega_2$ in the superconducting region, making clearer the regions with dominance of a SS or TS eigenvalues, as well as their dependence with system size and interactions.  We see that 
the critical interaction 
$U^*_{\rm{ss-ts}}(L,V/t)$ for a change of singlet/triplet dominance 
increases (in modulus) for larger system sizes as well as for larger (in modulus) inter-site interaction. A quantitative analysis of the transition line between the two different superconduting phases in the thermodynamic limit ($U^*_{\rm{ss-ts}}(L,V/t)$ for $L \rightarrow \infty$) is beyond the scope of this manuscript. It requires the analysis of much larger system sizes, which at the moment are numerically too expensive. It is worth remarking that even though a TS phase ``enlarges'' in the phase diagram for increasing system sizes, the existence of two different SS/TS orderings is still present in the thermodynamic limit. We can simply notice that in the region of singlet dominance,  e.g. for $V/t=-0.5$ and $U/t \sim -1 $, the gap $\Omega_2$ becomes larger as we increase the system size, showing that the singlet eigenvalue shall be dominant in the thermodynamic limit. A similar trend occurs in the region of triplet dominance,  e.g. for $V/t=-0.5$ and $U/t \sim -0.25 $, where for large enough system sizes we see a dominance of triplet eigenvalues ($L \sim 18$ sites), and further increasing the system size the gap inscreases as well, corroborating the triplet eigenvalue dominance in the thermodynamic limit.

\begin{figure}
 \includegraphics[scale=1.0,width=0.49\columnwidth]{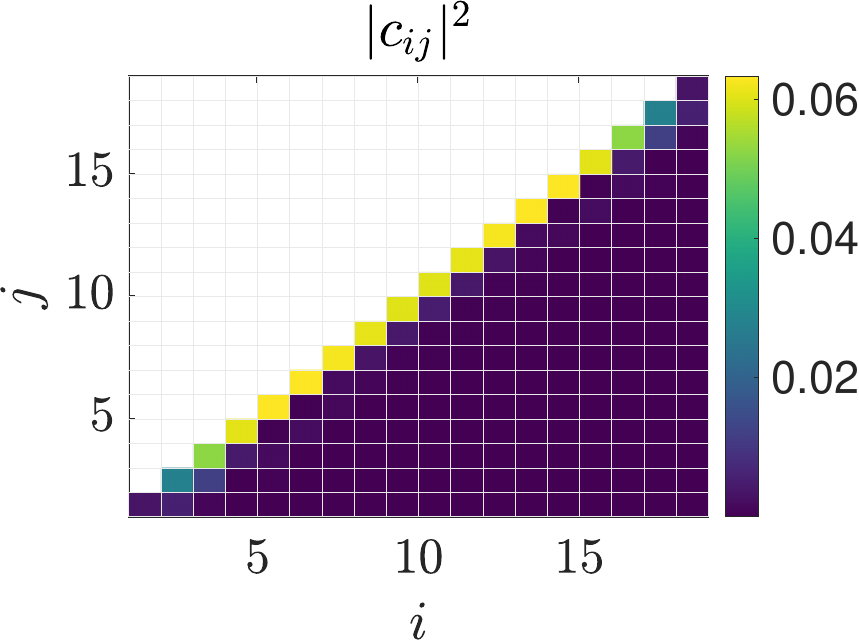}
 \includegraphics[scale=1.0,width=0.49\columnwidth]{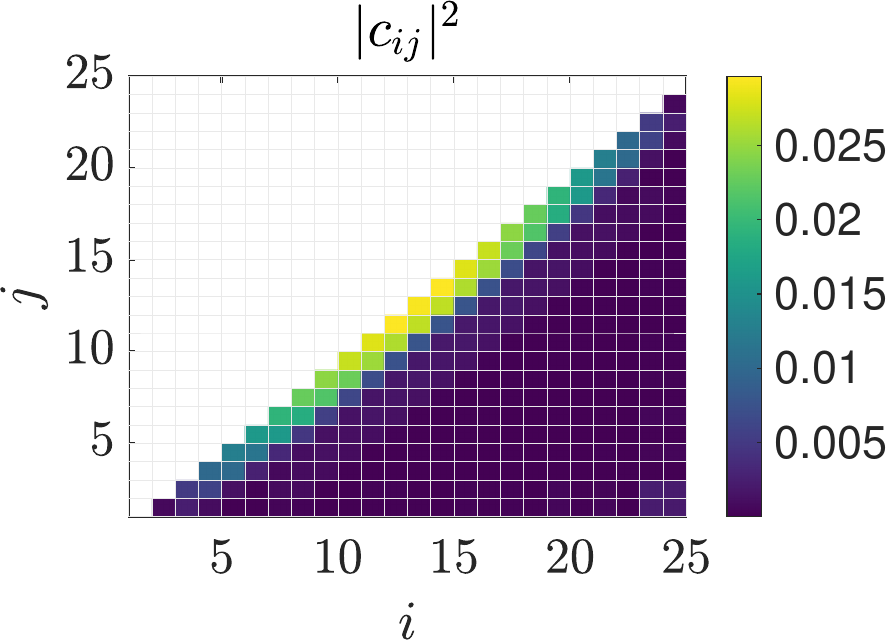}
\includegraphics[scale=1.0,width=0.49\columnwidth]{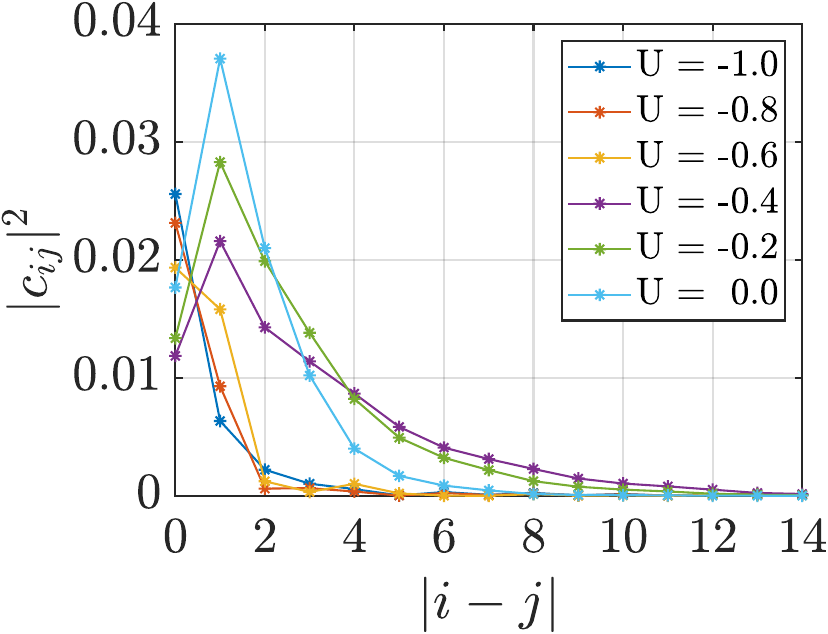}
\includegraphics[scale=1.0,width=0.49\columnwidth]{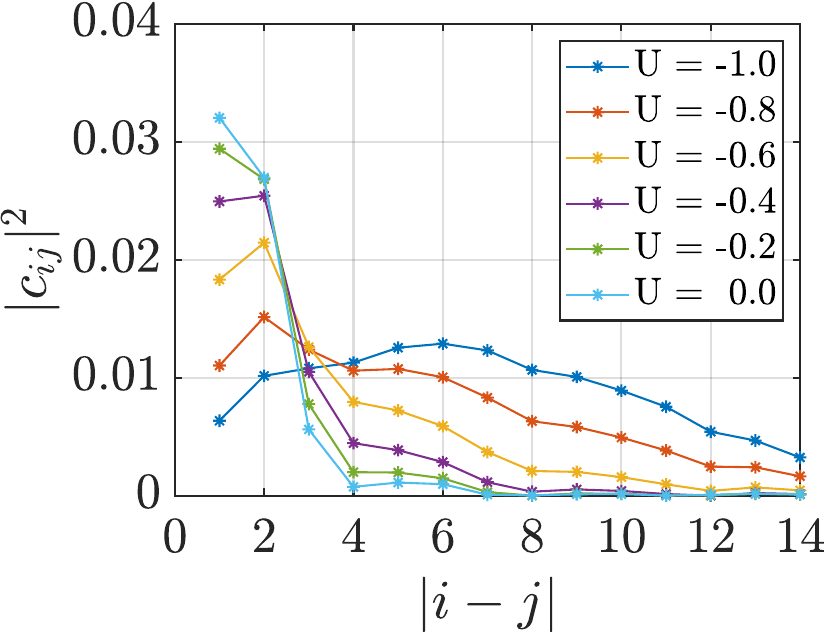}
\caption{ 
Coherence profiles $|c_{ij}|^2$ for the eigenvectors of the reduced density matrix. In the top panels we show the coherence profile for the dominant eigenvector in a system with 
\textbf{(top-left)} $L=18$ sites, $V/t=-0.5$, $U/t=-3.4$ and
\textbf{(top-right)} $L=24$ sites, $V/t=-0.6$,  $U/t=-0.2$, in the singlet and triplets subspaces respectively.
In the bottom panels we show the coherences for the dominant eigenvectors of the 
\textbf{(bottom-left)} singlet and \textbf{(bottom-right)} triplet subspaces, in a 
system with  $L=24$, $V/t=-0.6$ for varying on-site interactions.
In the bottom panels, the coefficients $c_{ij}$ are choosen close to the middle of the chain to minimize finite size effects and represent the bulk of the system (in the thermodynamic limit they will only depend on the distance $|i-j|$).}
\label{fig.dom.eigenvector}
\end{figure}

\textit{Dominant eigenvectors. } We perform a deeper analysis of the dominant eigenvalue of the reduced density matrix, studying their eigevector structure. In order to understand how the fermions are ordered within the eigenvector, we study its coherent superpositions. Specifically, we introduce a ``canonical'' basis in real space for the antisymmetric Hilbert space of two-fermions, in the singlet ($|s^0_{ij}\rangle$) and triplet ($|t^{[...]}_{ij}\rangle$) subspaces, as follows, 
\begin{eqnarray}
 |s_{0,ij}\rangle &=& \frac{(\ad{i \uparrow} \ad{j \downarrow} - \ad{i \downarrow} \ad{j \uparrow} )}{\sqrt{2}} \ket{vac},\\
 |t_{0,ij}\rangle &=& \frac{(\ad{i \uparrow} \ad{j \downarrow} + \ad{i \downarrow} \ad{j \uparrow} )}{\sqrt{2}} \ket{vac}, \\
 |t_{+(-),ij}\rangle &=& \ad{i\uparrow(\downarrow)} \ad{j \uparrow(\downarrow)}\ket{vac}.
\end{eqnarray}
The dominant eigevector $|D\rangle$ can always be decomposed in such a basis, $|D\rangle = \sum_{i,j} c_{ij} |s^0 (t^{[0,\pm]})_{ij} \rangle$, depending if it belongs to the singlet or triplet subspace.
 We show in Fig.\eqref{fig.dom.eigenvector} the coherence profiles for the dominant eigenvector along the superconducting region. In Fig.\eqref{fig.dom.eigenvector}-(top left) we show the case where the system belongs to a SS phase, with interacting values $U/t=-3$, $V/t=-0.6$ and the dominant eigenvector belonging to the singlet subspace. In this case the coherence profile $|c_{ij}|^2$ shows an almost uniform distribution along the chain for a fixed distance $|i-j|$ between the sites, displaying in this way a coherent superposition of fermionic pairs along all the chain of the system. Moreover, the coherence is maximum for fermion pairs at the same site ($i=j$), i.e., the fermions prefer a spatially local pairing ordering.
 In Fig.\eqref{fig.dom.eigenvector}-(top right) we show now the case where the system belongs to a TS phase, with interacting values $U/t=-0.2$, $V/t=-0.6$ and the dominant eigenvector belonging to the triplet subspace. We see a similar profile with, however, a predominance  of triplet pairs between nearest-neighbor sites (we recall that triplet pairs are forbidden to  occupy the same site, $i\neq j$).
 
 In Fig.\eqref{fig.dom.eigenvector}-(bottom pannels) we show the coherence of the dominant eigenvectors in the singlet and triplet subspaces, highlighting their dependence with the distance between the pairs.
  Interestingly, we notice that as we move from the SS phase towards the TS phase, the singlet pairs tend to 
spatially move apart from each other, while the triplet pairs tend to get closer together.
This indicates that the dominance of each pairing phase is related to the distance between the fermion pairs in the model - spatially closer (and coherent) pairings prompt a stronger superconducting ordering.

\subsection{$U\sim V $: Strong Coupling Regime ($U,V \gg t $)}
\label{subsec.u.sim.v}

Along the line with roughly equal couplings $U\sim V$ and in the strong coupling regime $U,V \gg t $ the entanglement gap displays a minimum, indicating in this way a possible phase transition in the model not yet discussed in the literature.  Since they persist along the strong coupling regime, one can better analyse the system within perturbation theory picture. 
Along this approach we observe that there are no macroscopic changes in the properties of the ground state, rather they follow at a few-body level (different types of local defects in the wavefunction). 
Let us analyse the two cases, repulsive $U,V>0$ and attractive $U,V<0$, separately.

\textit{Case $U,V>0$ (repulsive):} In the region  with $V \gtrsim U/2$ and considering the infinite coupling limit $t \rightarrow 0$, the ground state is characterized by macroscopic charge-density-wave configurations, such as $|2020...20\rangle$ or $|0202...02\rangle$. However, due to the open boundary conditions in the chain, these CDW configurations may present ``defects'', such as singly-occupied or empty nearest-neighbor sites. The energetic cost of the different defects depend on the strength of the couplings $V$ and $U$, and in this way one of them may prevail over the others depending on the ratio $V/U$. Specifically, the different CDW ground states and their corresponding defects 
are shown below:

$\bullet U \gtrsim  V \gtrsim U/2$ (singly occupied nearest-neighbor defects): The degenerate ground states and energy are given by,
\begin{eqnarray}
 |\rm{gs}\rangle_\ell &=& |\rm{CDW}^{[20]}\rangle_\ell \otimes |1 1\rangle \otimes 
 |\rm{CDW}^{[02]} \rangle_{L-\ell-2}, \nonumber \\
 \frac{E_{\rm{gs}}}{L U} &=& \frac{1}{2} + \frac{1}{L}(v-1),
\end{eqnarray}
where 
\begin{eqnarray}
 |\rm{CDW}^{[20]}\rangle_{\ell} &=& \otimes_{k=1}^{\ell/2} |20\rangle, \\
 |\rm{CDW}^{[02]}\rangle_{\ell} &=& \otimes_{k=1}^{\ell/2} |02\rangle 
\end{eqnarray}
with $\ell = 0,2,4,...,L$ indicating the sizes of the two possible charge-density-wave configurations in the degenerate subspace, $v=V/U$ and $E_{\rm gs}$ the ground state energy.

$\bullet V \gtrsim U$ (empty nearest-neighbor defects): In this case the degenerate ground states and energy are given by,
\begin{eqnarray}
 |\rm{gs}\rangle_\ell &=& |\rm{CDW}^{[20]}\rangle_\ell \otimes 
 |\rm{CDW}^{[02]} \rangle_{L-\ell}, \nonumber \\
 \frac{E_{\rm{gs}}}{L U} &=& \frac{1}{2},
\end{eqnarray}
with $\ell = 0,2,...,L$ indicating the defect position, i.e., the pair of empty $(\ell,\ell+1$) nearest-neighbor sites.

Therefore, despite the system is always characterized by a macroscopic CDW phase for both cases, at a few-body level the wavefunction has different properties. Considering a nonzero small hopping in the system $t \sim \epsilon$  shall not  change significantly this picture, leading only to second-order perturbative corrections to the wavefunction. We conclude in this way that the entanglement gap minima observed in Fig.\eqref{phase.diagram.q2} along these couplings are a consequence of the different types of few-body defects in the ground state wavefunction. We also remark that along this line the closure of entanglement gap follows between singlet eigenvalues (not shown), a different phenomenology as compared to the triplet-singlet supercondcuting phase transition dicussed in the previous section.

\textit{Case $U,V<0$ (atractive):} The analysis of the atractive case follows similarly to the previous one. The different defects and fluctuations, however, appear in the ground state from higher orders corrections in perturbation theory. Therefore, despite the entanglement gap may still capture these fluctuations are much weaker, in accordance with our numerical results, where the minimum observed in the quantifier is much smoother as compared to the repulsive case. Specifically, considering the infinite coupling limit the degenerate ground states are given by clusters of fully occupied sites, such as $|...022...220...\rangle$. Perturbative corrections correspond e.g. to hoppings of single fermions ($|20\rangle \rightarrow |11\rangle$) or doublons ($|20\rangle  \rightarrow |02\rangle$) at the edges of the clusters. The prevalence of these two ``defects'' depend on the ratio $V/U$, thus leading to different few-body quantum fluctuations in the wavefunction.

\section{Conclusions}
\label{sec.conclusions}

In this work we studied the ground state properties of the one-dimensional extended Hubbard model, composed of half-spin fermions, from the perspective of its \textit{particle} reduced density matrices. Focusing in the case of two-fermion reduced density matrices, we 
studied different facets of the quantum correlations and coherence on such states borrowing tools from Quantum Information and Entanglement theories. Specificaly, we analysed (i) the entanglement entropy of the reduced states, corresponding to the entanglement between $2$ fermions with the rest of $N-2$ fermions in the system; (ii) the irreducible two-body correlations contained in the cumulant matrix;  (iii) quantum coherences obtained from the off-diagonal elements of the reduced density matrix elements and (iv) the spectral structure and gap of the reduced density matrix.

In a general form, we obtained that all of the above quantifiers provide a qualitative view of the phase diagram of the model, showing peculiar behaviors such as discontinuities, maximum or minimum
values at the quantum phase transitions and are complementary to each other for a better description of the system properties. 
Interestingly, performing a finite-size scaling analysis of the quantifiers around the BOW related phase transitions, i.e., for a fixed $U/t=4$ and varying $V/t$ for different system sizes, we found that while the entanglement of particles $Q_2$, the coherence $C_2$ and the entanglement gap $\Omega_2$ have their maximum/minimum at the critical value $V^* \approx 2.22$ in the thermodynamic limit, the irredutible correlations $D_2$  have critical value closer to  $V^* \approx 2.02$. Comparing these results with the literature we tend to conclude that while $Q_2, C_2$ and $\Omega_2$ are most sensitive to the BOW-CDW phase transition, $D_2$ on the other hand is most related to the SDW-BOW phase transition.

We observed (numerically) that the two-fermion reduced density matrix has a $su(2)$ symmetry, thus spliting the Hilbert space into singlet and triplet subspaces. The overlap of the reduced matrix on such subspaces is intriguingly constant along all phase diagram, depending only on the number of sites $L$ in the system. In the thermodynamic limit the overlap onto all subspaces tend to be equal.

These symmetries opened interesting avenues for the investigation of the spectral properties of the reduced state. Focusing our analysis on the superconducting region of the phase diagram, 
we first observed the the dominant (largest) eigenvalues of the reduced matrix on the different subspaces cross at a critical interacting strength $U_{ss-ts}^*(L,V/t)$. Thus the dominat eigenvalue of the reduced state shifts from the singlet subspace to the triplet one, signaling different dominat pairing orderings for the fermionic particles.
Moreover, studying the structure of the dominant eigenvector, we showed that the dominance of a singlet or triplet eigenvalue in the spectrum of the reduced density matrix is related to the spatial distance between the fermionic pairs on the eigenvectors.
 Precisely, within the singlet subspace the fermion pairs tend to be spatially closer to each other when the singlet eigenvalue is dominant in the full spectrum. As one moves towards the TS phase these singlet pairs tend to move apart from each other, as well as decreasing the corresponding singlet eigenvalue. The same mechanism occurs for the triplet pairs and their eigenvalues.
 
 An interesting perspective for our work stands on delineating possible connections between our quantifiers with other approaches with more direct experimental access, such as optical conductivity and optical gap studies  \cite{jeckelmann02,Kancharla2001}. These could be experimentally probed by spectroscopy approaches, and similarly to our quantifiers they are also based on two-particle correlations. Nevertheless, a direct relation among them is not straightforward and would require a deeper analysis.
 
\section*{ACKNOWLEDGMENTS}
Financial support by the Brazilian agencies FAPEMIG (Fundação de Amparo à Pesquisa do Estado de Minas Gerais), CAPES (Coordenação de Aperfeiçoamento de Pessoal de Nível Superior),  CNPq (Conselho Nacional de Desenvolvimento Científico e Tecnológico) and INCT-IQ (National Institute of Science and Technology for Quantum Information).
F.I. acknowledges the financial support of the Brazilian funding agencies National Council
for Scientific and Technological Development—CNPq (Grant No. $308205$/$2019$-$7$) and FAPERJ (Grant No. E-$26$/$211.318$/$2019$).
 We also thanks to the ITensor library \cite{itensorLib} providing the basic functionality to construct our code 
which is available in the public repository \cite{myrepo} and can be freely used under its license.


\begin{thebibliography}{99}

\bibitem{Fraser2006} Fraser, G. (Ed.). (2006). \textit{The New Physics: For the Twenty-First Century}. Cambridge: Cambridge University Press. doi:10.1017/CBO9781139644228.

\bibitem{Xiao2013} Xiao-Gang Wen, \textit{Topological Order: From Long-Range Entangled Quantum Matter to a Unified Origin of Light and Electrons}, ISRN Condensed Matter Physics, vol. 2013, Article ID 198710, 20 pages, 2013. 

\bibitem{BeiZeng2018} Bei Zeng, Xie Chen, Duan-Lu Zhou, Xiao-Gang Wen, \textit{Quantum Information Meets Quantum Matter -- From Quantum Entanglement to Topological Phase in Many-Body Systems}, arXiv:1508.02595 (2018).

\bibitem{Osterloh2002} A. Osterloh, Luigi Amico, G. Falci and Rosario Fazio, \textit{Scaling of entanglement close to a quantum phase transition}. Nature 416, 608–610 (2002) doi:10.1038/416608a

\bibitem{amico08} L. Amico, R. Fazio, A. Osterloh and V. Vedral,
 Rev. Mod. Phys., 
\textbf{80}, 517 (2008).

\bibitem{ModiReview2012} Kavan Modi, Aharon Brodutch, Hugo Cable, Tomasz Paterek, and Vlatko Vedral, Rev. Mod. Phys. 84, 1655 (2012).


\bibitem{HorodeckiReview2009} Ryszard Horodecki, Paweł Horodecki, Michał Horodecki, and Karol Horodecki, Rev. Mod. Phys. 81, 865 (2009).

\bibitem{AlexanderReview2017} Alexander Streltsov, Gerardo Adesso, Martin B. Plenio, Rev. Mod. Phys. 89, 041003 (2017)

\bibitem{Li2008} Hui Li and F. D. M. Haldane,
Phys. Rev. Lett. 101, 010504 (2008).

\bibitem{Fidkowski2010} Lukasz Fidkowski,
Phys. Rev. Lett. 104, 130502 (2010).

\bibitem{Turner2011} Ari M. Turner, Frank Pollmann, and Erez Berg, Phys. Rev. B 83, 075102 (2011).


\bibitem{TRO2006a} Thiago R. de Oliveira, Gustavo Rigolin, Marcos C. de Oliveira and Eduardo Miranda, Phys. Rev. Lett. 97, 170401 (2006).

\bibitem{TRO2006b} Thiago R. de Oliveira, Gustavo Rigolin and Marcos C. de Oliveira, 
Phys. Rev. A 73, 010305(R).

\bibitem{Hofmann2014} Martin Hofmann, Andreas Osterloh and Otfried Gühne, Phys. Rev. B 89, 134101 (2014).

\bibitem{Konstantin2017} Konstantin V. Krutitsky, Andreas Osterloh, Ralf Schützhold,
Scientific Reports 7, 3634 (2017).


\bibitem{Antonio2020} Antônio C. Lourenço, Susane Calegari, Thiago O. Maciel, Tiago Debarba, Gabriel T. Landi and Eduardo I. Duzzioni, Phys. Rev. B \textbf{101}, 054431 (2020).


\bibitem{Iemini2016_concurrence} Fernando Iemini, Angelo Russomanno, Davide Rossini, Antonello Scardicchio and Rosario Fazio, Phys. Rev. B 94, 214206 (2016).

\bibitem{Lima2021} L. S. Lima, Eur. Phys. J. D \textbf{75}, 28 (2021), \url{https://doi.org/10.1140/epjd/s10053-021-00044-4}. 

\bibitem{Campbell2017_concurrence} S. Campbell, M.J.M Power and G. De Chiara, Eur. Phys. J. D \textbf{71}, 206 (2017). 

 \bibitem{Beggi2016} A. Beggi, F. Buscemi and P. Bordone,
Quantum Inf. Process. \textbf{15}, 3711–3743 (2016).

\bibitem{Iemini2015} Fernando Iemini, Thiago O. Maciel and Reinaldo O. Vianna,
Phys. Rev. B 92, 075423 (2015).

\bibitem{Vianna2016} A. L. Malvezzi, G. Karpat, B. Çakmak, F. F. Fanchini, T. Debarba and R. O. Vianna,
Phys. Rev. B 93, 184428 (2016)

\bibitem{Chiara2018} Gabriele De Chiara and Anna Sanpera, Rep. Prog. Phys. \textbf{81} 074002 (2018). 

\bibitem{bagrov2020} Andrey A. Bagrov, Mikhail Danilov, Sergey Brener, Malte Harland, Alexander I. Lichtenstein and Mikhail I. Katsnelson, 
Sci. Rep. \textbf{10}, 20470 (2020).

\bibitem{hubbard63} J. Hubbard, Proc. R. Soc. Lond. A
\textbf{276}, 238-257 (1963).




\bibitem{Masudul2007} Masudul Haque, Oleksandr Zozulya and Kareljan Schoutens,
Phys. Rev. Lett. 98, 060401 (2007).

\bibitem{Lukas2017} Lukas Rammelmüller, William J. Porter, Jens Braun and Joaquín E. Drut,
Phys. Rev. A 96, 033635 (2017).


\bibitem{Zhao2010} Zhao Liu and Heng Fan,
Phys. Rev. A 81, 062302 (2010).

\bibitem{Herdman2015} C. M. Herdman, A. Del Maestro,
Phys. Rev. B 91, 184507 (2015)

\bibitem{Masudul2009} Masudul Haque, O. S. Zozulya and K Schoutens,
J. Phys. A: Math. Theor. 42, 504012 (2009).

\bibitem{Hatem2017} Hatem Barghathi, Emanuel Casiano-Diaz and Adrian Del Maestro,
Journal of Statistical Mechanics: Theory and Experiment, 
8, 083108 (2017).

\bibitem{Alexandradinata2011} A. Alexandradinata, Taylor L. Hughes and B. Andrei Bernevig, Phys. Rev. B 84, 195103 (2011)

 \bibitem{Tullio2019} M. Di Tullio, R. Rossignoli, M. Cerezo and N. Gigena, Phys. Rev. A \textbf{100}, 062104 (2019).

\bibitem{Sterdyniak2011} A. Sterdyniak, N. Regnault and B. A. Bernevig,
Phys. Rev. Lett 106, 100405 (2011).

\bibitem{Rex2014} Rex Lundgren, Jonathan Blair, Martin Greiter, Andreas Läuchli, Gregory A. Fiete and Ronny Thomale,
Phys. Rev. Lett 113, 256404 (2014).



\bibitem{Jian2004} Shi-Jian Gu, Shu-Sa Deng, You-Quan Li and Hai-Qing Lin,
Phys. Rev. Lett. 93, 086402 (2004).

\bibitem{Deng2006} Shu-Sa Deng, Shi-Jian Gu, and Hai-Qing Lin, Phys. Rev. B 74, 045103 (2006).

\bibitem{Yang2008} Yang Zhen, Ning Wen-Qiang,
Vol. 25, No. 1 (2008) 31.

\bibitem{Mund2009} C. Mund, O. Legeza and R. M. Noack, 
Phys. Rev. B 79, 245130 (2009).

\bibitem{Liu2011} Liu Guang-Hua and Wang Chun-Hai,
Commun. Theor. Phys. 55, 702–708 (2011).

 \bibitem{Barbiero2017} L. Barbiero, S. Fazzini, A. Montorsi, Eur. Phys. J. Spec. Top. \textbf{226}, 2697 (2017). 

\bibitem{Gigena2013} N. Gigena and R. Rossignoli, 
Phys. Rev. A \textbf{92}, 042326 (2015).

 \bibitem{Chung2021} M.H. Chung, J. Korean Phys. Soc. \textbf{78}, 700–705 (2021).

 \bibitem{Rausch2020} Roman Rausch and Matthias Peschke,
 New J. Phys. \textbf{22}, 073051 (2020).

 \bibitem{Pandey2017} Bradraj Pandey and Swapan K. Pati, 
 Phys. Rev. B \textbf{95}, 085105 (2017).
 
 

\bibitem{iemini13b} F. Iemini,  and R. O. Vianna, Phys. A. Rev
\textbf{87}, 022327 (2013).

\bibitem{iemini13a} F. Iemini, T. O. Maciel,
 T. Debarba, and R. O. Vianna, Quantum Inf. Process.,
\textbf{12}, 733 (2013).

\bibitem{iemini14} F. Iemini, T. Debarba, and R. O. Vianna, Phys. Rev. A, {\bf 89}, 032324 (2014).

\bibitem{balachandran1} A. P. Balachandran, T. R. Govindarajan,
 A. R. de Queiroz and A. F. Reyes-Lega,
Phys. Rev. Lett. {\bf 110}, 080503 (2013).

\bibitem{balachandran2} A. P. Balachandran,
 T. R. Govindarajan, A. R. de Queiroz and A. F. Reyes-Lega,
Phys. Rev. A {\bf 88}, 022301 (2013).

\bibitem{ghirardi02} G.C. Ghirardi, L. Marinatto and T. Weber,
J. Stat. Phys. {\bf 108}, 49 (2002).

\bibitem{ghirardi04} G.C Ghirardi and L. Marinatto,
 Phys. Rev. A {\bf 70}, 012109 (2004).

\bibitem{schliemann01a} J. Schliemann, D. Loss, and A. H. MacDonald,
Phys. Rev B {\bf 63}, 085311 (2001).

\bibitem{schliemann01b} J. Schliemann, J. I. Cirac, M. Kus, M. Lewenstein and D. Loss,
Phys. Rev. A {\bf 64}, 022303 (2001).

\bibitem{eckert02} K. Eckert, J. Schliemann, D. Bruss and M. Lewenstein,
Ann. Phys. {\bf 299}, 88 (2002).

\bibitem{li01} Y. S. Li, B. Zeng, X. S. Liu and G. L. Long,
Phys. Rev. A {\bf 64}, 054302 (2001).

\bibitem{barnum04} H. Barnum, E. Knill,
 G. Ortiz, R. Somma, and L. Viola, Phys.
Rev. Lett. \textbf{ 92}, 107902 (2004).

\bibitem{somma04} Rolando Somma, Gerardo Ortiz, Howard Barnum, Emanuel Knill 
and Lorenza Viola, Physic. Rev. A \textbf{70}, 042311 (2004).

\bibitem{paskauskas01} R. Paskauskas and L. You, Phys. Rev. A
\textbf{64}, 042310 (2001).

\bibitem{plastino09} A. R. Plastino, D. Manzano and J. S. Dehesa, 
Europhys. Lett \textbf{86}, 20005 (2009).

\bibitem{zander10} C. Zander and A. R. Plastino,
Phys. Rev. A {\bf 81}, 062128 (2010).

\bibitem{Debarba2017} Tiago Debarba, Reinaldo O. Vianna, and Fernando Iemini,  Phys. Rev. A 95, 022325 (2017)

\bibitem{ding2020} Lexin Ding and Christian Schilling,
J. Chem. Theory Comput. \textbf{16} (7), 4159-4175 (2020).

\bibitem{ding2021} Lexin Ding, Sam Mardazad, Sreetama Das, Szilárd Szalay, Ulrich Schollwöck, Zoltán Zimborás, and Christian Schilling,
J. Chem. Theory Comput. \textbf{17} (1), 79-95 (2021).

\bibitem{Gigena2017} N. Gigena and R. Rossignoli,
Phys. Rev. A \textbf{ 95}, 062320 (2017).

\bibitem{Tullio2018} M. Di Tullio, N. Gigena and R. Rossignoli,
Phys. Rev. A \textbf{97}, 062109 (2018).

\bibitem{Gigena2020a} N. Gigena, M. Di Tullio and R. Rossignoli,
Phys. Rev. A \textbf{102}, 042410 (2020).

\bibitem{Benatti2020} F.Benatti, R.Floreanini, F.Franchini and U.Marzolino, Physics Reports \textbf{878}, 1-27 (2020).

 \bibitem{Benavoli2021} Alessio Benavoli, Alessandro Facchini and Marco Zaffalon, arXiv:2105.04336 (2021).

\bibitem{Gigena2020b} N. Gigena, M. Di Tullio, R. Rossignoli, arXiv:$2012.13785$ (2020).

\bibitem{kubo62} Ryogo Kubo, \textit{Generalized Cumulant Expansion Method}, Journal of the Physical Society of Japan, Vol.17, No. 7, 1100 (1962).

\bibitem{Werner1999} Werner Kutzelnigg,
J. Chem. Phys. 110, 2800 (1999).


\bibitem{Shi2003} Yu Shi, Phys. Lett. A 309, 254-261 (2003).

\bibitem{Patrycja2019}Patrycja Łydżba, Tomasz Sowiński, 
	arXiv:1911.04188 (2019).

\bibitem{verstraete2008matrix} F. Verstraete, V. Murg and J. I. Cirac,  Advances in Physics \textbf{57}, 143–224 (2008).

\bibitem{nielsen2005fermionic} M. Nielsen,
The fermionic canonical commutation relations and the jordan-wigner transform, $2005$.

\bibitem{Baumgratz2014} T. Baumgratz, M. Cramer, M. B. Plenio, 
Phys. Rev. Lett. 113, 140401 (2014)


\bibitem{solymon}  J. S\'olymon, Adv. Phys., {\bf 28}, 201-303 (1979).

\bibitem{DMRG} U. Schollw\"ock, Ann. Phys., \textbf{326}, 96-192 (2011).

\bibitem{ALPS} B. Bauer, L.D. Carr, H.G Evertz, A. Feiguin, J. Freire, S. Fuchs, L. Gamper, J. Gukelberger, E. Gull, S. Guertler, A. Hehn, R. Igarashi, S.V. Isakov, 
D. Koop, P.N. Ma, P. Mates, H. Matsuo, O. Parcollet, G. Paw{\l}owski, J.D. Picon, L. Pollet, E. Santos, V.W. Scarola, U. Schollw{\"o}ck, C. Silva, B. Surer, S. Todo, 
S. Trebst, M. Troyer, M.L. Wall, P. Werner,  and S. Wessel, J. Stat. Mech, {\bf 2011}, P05001 (2011).

\bibitem{lin00} H. Q. Lin, D. K. Campbell and R. T. Clay, Chinese Journal of Physics
\textbf{38}, 1 (2000).

\bibitem{nakamura00} M. Nakamura, Phys. Rev. B 
\textbf{61}, 16377 (2000).

\bibitem{sengupta02} P. Sengupta, A. W. Sandvik and D. K. Campbell, Phys. Rev. B 
\textbf{65}, 155113 (2002).

\bibitem{zhang04} Y. Z. Zhang, Phys. Rev. Lett. 
\textbf{92}, 246404 (2004).

\bibitem{dalmonte14} M. D., J. Carrasquilla, L. Taddia, E. Ercolessi and M. Rigol, arXiv:1412.5624v1

\bibitem{jeckelmann02} E. Jeckelmann, Phys. Rev. Lett. 
\textbf{89}, 236401 (2002).

\bibitem{Kancharla2001} S. S. Kancharla and C. J. Bolech, 
Phys. Rev. B \textbf{64}, 085119 (2001).

\bibitem{ejima07} Satoshi Ejima and S. Nishimoto, Phys. Rev. Lett.
\textbf{99}, 216403 (2007).

\bibitem{sandvik04} A. W. Sandvik, L. Balents and D. K. Campbell, Phys. Rev. Lett.
\textbf{92}, 236401 (2004).

\bibitem{Shinjo2019}
 Kazuya Shinjo, Kakeru Sasaki, Satoru Hase, Shigetoshi Sota, Satoshi Ejima, Seiji Yunoki  and Takami Tohyama,
 Journal of the Physical Society of Japan,
\textbf{88},065001 (2019), \url{https://doi.org/10.7566/JPSJ.88.065001}

\bibitem{Kazuya2020} Kazuya Shinjo, Shigetoshi Sota, Seiji Yunoki and Takami Tohyama, Phys. Rev. B 101, 195136 (2020).

\bibitem{DaiWei2021} Dai-Wei Qu, Bin-Bin Chen, Hong-Chen Jiang, Yao Wang and Wei Li, arXiv:2110.00564 (2021).


\bibitem{plastino2009} A. R. Plastino and D. Manzano and J. S. Dehesa, 
Europhysics Letters \textbf{86}, 20005 (2009)
\bibitem{plastino2016} Majtey, A. P. and Bouvrie, P. A. and Vald\'es-Hern\'andez, 
A. and Plastino, A. R.Phys. Rev. A \textbf{93}, 032335 (2016)

\bibitem{zanardi02} P. Zanardi, Phys. Rev. A 
\textbf{65}, 042101 (2002).

\bibitem{wiseman03} H. M. Wiseman and J. A. Vaccaro, Phys.
 Rev. Lett. \textbf{91}, 097902 (2003).

\bibitem{banulus07} M.C. Banuls, J. I. Cirac and M. M. Wolf,
Phys. Rev. A {\bf 76}, 022311 (2007).

\bibitem{benatti10} F. Benatti, R. Floreanini and U. Marzolino, Ann. Phys.
\textbf{325}, 924 (2010).

\bibitem{grabowski11} J. Grabowski, M. Kus and G. Marmo,
J. Phys. A: Math. Theor. {\bf44}, 175302 (2011). 

\bibitem{grabowski12} J. Grabowski, M. Kus and G. Marmo,
J. Phys. A: Math. Theor. {\bf 45}, 105301 (2012).

\bibitem{tsutsui} T. Sasaki, T. Ichikawa and I. Tsutsui,
Phys. Rev. A {\bf 83}, 012113 (2011).

\bibitem{reusch15} A. Reusch, J. Sperling  and W. Vogel,
 Phys. Rev. A \textbf{91} 042324, (2015).
 
\bibitem{dft89} R. G. Parr and W. Yang,
 Density Matrices: The N-representability of reduced
 density matrices. In: Density-Functional Theory of Atoms
 and Molecules, The International Series of Monographs
 on Chemistry - 16, Oxford Science Publications. 1989.

\bibitem{giamarchibook} T. Giamarchi, \textit{Quantum Physics in 
One Dimension} (Oxford University Press, New York, 2004).

\bibitem{mund09} C. Mund, \"{O}. Legeza and R. M. Noack, Phys. Rev. B 
\textbf{79},  245130  (2009).

\bibitem{jian04} Shi-Jian Gu, Shu-Sa Deng, You-Quan Li and Hai-Qing Lin, Phys. Rev. Lett.
\textbf{93}, 086402 (2004).

\bibitem{sa06} Shu-Sa Deng, Shi-Jian Gu and Hai-Qing Lin, Phys. Rev. B 
\textbf{74}, 045103 (2006).

\bibitem{essler2005} Essler, F., Frahm, H., Göhmann, F., Klümper, A., \& Korepin, V. (2005). The One-Dimensional Hubbard Model. Cambridge: Cambridge University Press. doi:10.1017/CBO9780511534843

\bibitem{itensorLib}
Matthew Fishman, Steven R. White and E. Miles Stoudenmire, \textit{
 The \mbox{ITensor} Software Library for Tensor Network Calculations},
 arXiv.2007.14822 (2020).
 
\bibitem{myrepo}
\url{https://github.com/diegobragaferreira/itensor-hubbardmodel-2particle}

\end{thebibliography}
\end{document}